\documentstyle[12pt,a41,epsfig,wrapfig,rotating]{article}

\begin{document}
\newcommand{\ds}{\displaystyle}
\newcommand{\la}[1]{\label{#1}}
\newcommand{\re}[1]{\ (\ref{#1})}
\newcommand{\nn}{\nonumber}
\newcommand{\be}{\begin{equation}}
\newcommand{\ee}{\end{equation}}
\newcommand{\ba}{\begin{eqnarray}}
\newcommand{\ea}{\end{eqnarray}}
\newcommand{\baz}{\begin{eqnarray*}}
\newcommand{\eaz}{\end{eqnarray*}}
\newcommand{\ct}[1]{${\cite{#1}}$}
\newcommand{\ctt}[2]{${\cite{#1}-\cite{#2}}$}
\newcommand{\bi}[1]{\bibitem{#1}}
\newcommand{\ep}{\varepsilon}
\newcommand{\epm}{\varepsilon_{\mu\nu\lambda\sigma}}
\newcommand{\epa}{\varepsilon_{\mu\alpha\nu\beta}}
\newcommand{\epb}{\varepsilon_{\mu\nu\alpha\beta}}
\newcommand{\Sa}{S_{\mu\alpha\nu\beta}}
\newcommand{\gb}{\gamma_\beta}
\newcommand{\g}{\gamma_5}
\newcommand{\lc}{\lambda^c}
\newcommand{\bq}{\bar q}
\hyphenation{Wand-zura}

\begin{flushleft}
DESY 98-181 \hfill {\tt hep-ph/9812478}\\
December 1998
\end{flushleft}
\vspace*{4cm}

\begin{center}
{\LARGE \bf Target Mass Corrections for}

\vspace{3mm}
{\LARGE \bf Polarized Structure Functions}

\vspace{4mm}
{\LARGE \bf and New Sum Rules}

\vspace*{2cm}
\large
{Johannes Bl\"umlein and Avtandil Tkabladze\footnote{
Alexander von Humboldt Fellow}}\\

\vspace*{1.cm}
\normalsize
{\it DESY Zeuthen, D-15738 Zeuthen, Germany}\\

\vspace{\fill}

\end{center}

\begin{abstract}
\noindent
The target mass corrections are calculated for all structure functions of
neutral and charged current deep inelastic scattering in lowest order in 
the coupling constant. Representations of the correction to the twist--2 
and twist--3 contributions are derived both in Mellin-$n$ and $x$-space. 
The impact of the target mass corrections on the general relations between
the twist--2 and twist--3 parts of the structure functions is studied and
three new relations between the twist--3 contributions are derived.
\end{abstract}

\vspace{\fill}

\newpage
\section{Introduction}
\setcounter{equation}{0}

\vspace{2mm}
\noindent
Deep inelastic lepton--nucleon scattering provides one of the cleanest
possibilities to study the nucleon structure at short distances. The
light--cone expansion~\cite{LC} proved to be one of the most powerful
techniques in the Bjorken limit, $Q^2, P.q \rightarrow \infty,
x = Q^2/2 P.q =~{\sf const.}$, to derive the structure of the scattering 
cross sections, relations between the structure functions and their 
scaling violations.

The early unpolarized deep inelastic scattering experiments~\cite{SLAC}
and most of the experiments, in which polarized lepton scattering off a 
polarized target~\cite{POLE, G2EXP} has been studied so far, operated in 
the range of lower values of $Q^2$. 
In this domain nucleon--mass corrections cannot be neglected.
The target mass effects of $O((M^2/Q^2)^k)$ form one contribution to the 
power corrections. Unlike the case for dynamical higher twist 
effects~[5--7] the target mass corrections can be calculated in closed
form in all orders in $M^2/Q^2$ for deep inelastic structure functions. 
Because the product of a mass factor $M^k$ and a {\it genuine} twist--$l$ 
operator forms an operator of twist $\tau = k+l$ the individual 
terms in the Taylor--expansion in $M^2/Q^2$ of the target mass corrections
mix with the dynamical higher twist operators under renormalization.
Therefore a fully consistent description would require a common treatment
of both contributions. Little is known so far on the relative strength of
dynamical higher twist operators and their scaling violations. Before
one may try to pursue a common treatment various conceptional problems
concerning higher twist operators have first to be solved. Due to this we 
limit the present analysis to a systematic study of the target mass 
corrections for all deep inelastic structure functions
extending earlier investigations~\cite{GP,PR}. 

Two methods were proposed in the literature for the evaluation of the
target mass corrections. Nachtmann~\cite{Nachtmann} translated the usual
power--series expansion into an expansion of operators of definite spin.
The problem may be solved by applying the representation--theory of the
Lorentz group. A second method is due to Georgi and Politzer~\cite{GP},
see also~\cite{XI},
in which the individual nucleon mass terms are collected and resummed.
In the domain of large values of $x$,  $x \sim 1$, these methods lead
to different results for the operators of lowest twist. This is the 
region, however, in which also the dynamical higher twist terms of any 
order contribute, which have to be included into the analysis in this
kinematic domain as well.

We will apply the method of Ref.~\cite{GP} to all polarized structure
functions.\footnote{The target mass corrections for $g_1$ and $g_2$ were 
derived in~\cite{WA,MU} and to the polarized Bjorken sum rule
in~\cite{KU} following Ref.~\cite{Nachtmann}.} The contributions due to 
the twist--2 
and twist--3 operators are studied individually. Both the corrections for
the Mellin moments of the different structure functions and the inverse 
Mellin transform to Bjorken-$x$ space, which can be derived analytically,
are provided. As will be shown, the complete resummation of the different
series in $(M^2/Q^2)^k$ is required to avoid artificially large terms in 
the region $x \rightarrow 1$.

A second goal of our investigation is to study the effect of the target 
mass corrections on the integral relations between the different
polarized structure functions in lowest order in the coupling constant.
As was shown in a previous analysis~\cite{BK} the twist--2 contributions
to the polarized structure functions are connected by three (integral) 
relations, the Dicus--relation~\cite{DIC}, the Wandzura--Wilczek 
relation~\cite{WW} and a new relation given in Ref.~\cite{BK}. A general 
relation which holds for all spins $n$ could also be derived for the 
valence part of the twist--3 contributions to the structure functions
$g_2$ and $g_3$. As in Ref.~\cite{BK} nucleon mass effects were 
disregarded, except of those implied by the kinematics in the Born cross
sections, 
twist--3 contributions to the structure functions $g_1, g_4$ and $g_5$ 
were not obtained. This picture has to be regarded as partly incomplete, 
since nucleon mass effects have either to be accounted for thoroughly or 
to be neglected at all. In the latter case, however, the scattering cross
sections for longitudinally polarized nucleons would even not contain the
structure functions $g_2$ and $g_3$ and therefore not allow to obtain two
of the twist--2 relations, namely the Wandzura--Wilczek relation and the 
new twist--2 relation of Ref.~\cite{BK}. On the other hand, a rather 
asymmetric picture is yet obtained for relations in the twist--3 sector. 
Particularly, there is no relation yet for the complete twist--3 term 
contained in $g_2$ and a potential other structure function. As will be 
shown below the {\it complete} inclusion of the target mass corrections 
solves this problem since all polarized structure functions obtain, 
besides the twist--2 contributions, by virtue of these corrections also 
twist--3 terms. Similar to the case of the twist--2 contributions 
observed before three {\it new} relations are implied for the twist--3 
parts of the structure functions in lowest order in the coupling constant.
These relations provide the possibility of a thorough test of the twist--3 
structure of the nucleon.

The paper is organized as follows. In section~2 the cross sections are 
summarized for neutral and charged current deep inelastic scattering
allowing also for contributions due to current non--conservation. In
section~3 the structure of the forward Compton amplitude is derived and 
section~4 provides the general expressions of the operator product 
expansion in the presence of target mass corrections. The nucleon matrix 
elements are evaluated in section~5. The relation of the moments of the 
twist--2 and twist--3 contributions to the different polarized structure 
functions are given in section~6. In section~7 the $x$--space 
representations for the twist--2 and twist--3 contributions of the target 
mass corrections to the polarized structure functions are provided. 
Section~8 deals with the effect of the target mass corrections on the 
twist--2 relations between the polarized structure functions in the 
massless limit. Three new relations are derived between the twist--3 
contributions of the polarized structure functions in section~9. There
also other twist--3 relations are discussed. Section~10 contains the
conclusions and an appendix deals with the quark--mass corrections in the 
case of the Wandzura--Wilczek relation.
\section{The Scattering Cross Section}

\vspace{2mm}
\noindent
The differential Born cross section for polarized lepton--polarized
nucleon scattering is given by
\be
\label{e1}
\frac{d^3 \sigma}{dx dy d\phi} = \frac{y \alpha^2}
{Q^4} \sum_{i} \eta_i(Q^2) L^{\mu\nu}_i W_i^{\mu\nu}~.
\ee
Here the index $i$ denotes the different current combinations,
i.e. $i = |\gamma|^2, |\gamma Z|, |Z|^2$ for neutral and 
$i = |W^{\pm}|^2$ for charged current interactions. $\phi$ is the 
azimuthal angle of the final-state lepton, $x = Q^2/(2 P.q) \equiv
Q^2/(2 \nu)$ and 
$y =  P.q/k.P$ are the Bjorken variables, where $q = k - k'$ the four 
momentum transfer to the hadronic vertex and $Q^2 = - q^2$. $P, k$ and 
$k'$ are the proton--, initial-- and final state lepton four--momenta,
respectively. The factors $\eta_i(Q^2)$ denote the ratios of the
corresponding propagator terms to the photon propagator squared,
\begin{eqnarray}
\label{etas}
\eta^{|\gamma|^2}(Q^2)  &=& 1 ,\nonumber\\
\eta^{|\gamma Z|}(Q^2)  &=& \frac{G_FM_Z^2}
{2\sqrt{2}\pi\alpha}\frac{Q^2}{Q^2+M_Z^2},
\nonumber\\
\eta^{|Z|^2}(Q^2)       &=& (\eta^{|\gamma Z|})^2(Q^2), \nonumber\\
\eta^{|W^{\pm}|^2}(Q^2) &=& \frac{1}{2}
\left( \frac{G_FM_W^2}{4\pi\alpha} \frac{Q^2}{Q^2+M_W^2} \right)^2~.
\end{eqnarray}
$\alpha$ is the fine structure constant, $G_F$ the Fermi
constant and $M_Z$ and $M_W$ are the $Z$ and $W$~boson masses.

The leptonic tensor has the following form
\be
\label{e2}
L_{\mu\nu}^i=\sum_{\lambda^\prime}\left[\bar
u(k^\prime,\lambda^\prime)\gamma_\mu(g_V^{i_1}+g_A^{i_1}
\gamma_5)u(k,\lambda)\right]^{\dag} \bar
u(k^\prime,\lambda^\prime)\gamma_\nu(g_V^{i_2}
+g_A^{i_2}\gamma_5)u(k,\lambda)~.
\ee
Here, $\lambda$  denotes the helicity of the initial state lepton. The 
indices $i_1$ and $i_2$ refer to the currents forming the combinations 
$i$ in Eq.~(\ref{e1}), and the vector and axial vector couplings read
\begin{equation}
\label{ac}
\begin{array}{lcllcl}
g_V^{\gamma} &=& 1, & g_A^{\gamma} &=&~~0~, \\
g_V^Z&=&\ds
\frac{1}{2}-2 Q_I\sin^2\theta_W,& g_A^Z &=&{ \ds -\frac{1}{2}~,} \\
g_V^{W^-} &=& 1, &g_A^{W^-} &=& -1~.
\end{array}
\end{equation}
\renewcommand{\arraystretch}{2}

\noindent
Here $Q_I$ denotes the charge of the initial--state lepton. $\theta_W$ is
the weak mixing angle. 

The hadronic tensor is given by
\ba
W_{\mu\nu}^{i}=\frac{1}{4\pi}\int d^4x e^{iqx}
\langle PS\mid[{J_\mu^{i_1}(x)}^\dagger ,J_\nu^{i_2}(0)]\mid PS\rangle~.
\ea
$S$ denotes the four--vector of the nucleon spin with $S\cdot P=0$. In 
the following we normalize $S^2 = -M^2$, where $M$ is the nucleon mass.
In the  framework of the quark--parton model the currents $J_{\mu}^j$ are
given by
\ba
\label{eqCUR1}
J_\mu^j(x)=\sum_{f,f'}\bar q_f'(x)\gamma_\mu(g_V^{j,f}+g_A^{j,f}
\gamma_5)q_f(x)U_{ff^\prime}~,
\ea
where  $g_{V,A}^{j,f}$ are the electroweak couplings of the quark labeled
by $f$. For charged current interactions $U_{ff^\prime}$ denotes the
Cabibbo-Kobayashi-Maskawa matrix and $g_V=1, {\ } g_A=-1$, whereas for 
neutral current interactions  $U_{ff^\prime} = \delta_{ff^\prime}$, 
$g_V^q = e_q, g_A = 0$ for $\gamma$, and
\be
\label{e111}
\begin{array}{lcrlcrl}
g_V^q &=& \ds
\frac{1}{2}-\frac{4}{3}\sin^2\theta_W,
& g_A^q &=& \ds
-\frac{1}{2}, &~~{\rm for}~~q=u,c,t \\\
g_V^q &=&\ds
-\frac{1}{2}+\frac{2}{3}\sin^2\theta_W,
& g_A^q &=& \ds
\frac{1}{2}, &~~{\rm for}~~ q=d,s,b
\end{array}
\ee
\renewcommand{\arraystretch}{1.5}

\noindent
for $Z$--boson exchange. 

The hadronic tensor is constructed requiring Lorentz and time--reversal 
invariance. The general structure of the hadronic tensor is
\begin{eqnarray}
W_{\mu\nu} =&& \left(-g_{\mu\nu}+\frac{q_\mu q_\nu}{q^2}\right)F_1(x,Q^2)
  + \frac{\hat P_\mu\hat P_\nu}{P\cdot q} F_2(x,Q^2)
-i\varepsilon_{\mu\nu\lambda\sigma}\frac{q^\lambda P^\sigma}{2P\cdot q} 
F_3(x,Q^2)
\nonumber \\
&&+\frac{q_\mu q_\nu}{P\cdot q} F_4(x,Q^2)
+\frac{(p_\mu q_\nu+p_\nu q_\mu)}{2 P\cdot q} F_5(x,Q^2) \nonumber \\
&& +i\varepsilon_{\mu\nu\lambda\sigma}\frac{q^\lambda S^\sigma}{P\cdot q} 
g_1(x,Q^2) 
+i\varepsilon_{\mu\nu\lambda\sigma}
\frac{q^\lambda (P\cdot q S^\sigma-S\cdot q P^\sigma)}{(P\cdot q)^2} 
g_2(x,Q^2) \nonumber \\
&&+\left[\frac{\hat P_\mu\hat S_\nu+\hat S_\mu\hat P_\mu}{2} -
S\cdot q\frac{\hat P_\mu\hat P_\nu}{P\cdot q}\right] \frac{g_3(x,Q^2)}
{P\cdot q}\nonumber \\
&&+S\cdot q\frac{\hat P_\mu\hat P_\nu}{(P\cdot q)^2} g_4(x,Q^2) +
(-g_{\mu\nu}+\frac{q_\mu q_\nu}{q^2})\frac{(S\cdot q)}{P\cdot q} g_5(x,Q^2),
\nonumber\\
&& + i\varepsilon_{\mu\nu\lambda\sigma} \frac{P_\sigma S_\lambda}{P\cdot q} 
g_6(x,Q^2) +
S\cdot q\frac{q_\mu q_\nu}{(P\cdot q)^2} g_7(x,Q^2) \nonumber \\
&&+\frac{(p_\mu q_\nu+q_\mu p_\nu) S\cdot q}{2 (P\cdot q)^2} g_8(x,Q^2)
+\frac{S_\mu q_\nu+S_\nu q_\mu}{2 P\cdot q} g_9(x,Q^2)~,
\label{had1}
\end{eqnarray}
where 
\begin{eqnarray}
\hat P_\mu = P_\mu-\frac{P\cdot q}{q^2}q_\mu~,~~~~~~~~~~~~~~~~~
\hat S_\mu = S_\mu-\frac{S\cdot q}{q^2}q_\mu~.
\nonumber
\end{eqnarray}
Here the current indices were suppressed. In general the hadronic tensor 
is determined by five unpolarized structure functions $F_i$ and nine 
polarized structure functions $g_i$. Due to the structure of the 
electroweak couplings $g_{V,A}^q$ given above and due to current
conservation, for photon exchange only the structure functions $F_{1,2}$
and $g_{1,2}$ contribute. For the weak currents in general all structure 
functions given above are present. The notation for the structure 
functions $F_i$, $g_1$ and $g_2$ is widely unique in the literature. 
However, different notations are used for the structure functions 
$\left. g_i\right|_{i=3}^9$. We follow the  definitions of 
Ref.~\cite{BK} for the structure functions $\left. g_i\right|_{i=3}^5$.
For the current non--conserving structure functions our definitions 
coincide with those introduced in Ref.~\cite{Ji}, cf. also \cite{FRANKF}.

From Eqs.~(\ref{e1},\ref{e2}) and (\ref{had1}) one obtains the 
differential scattering cross sections of a lepton with helicity 
$\lambda$ off a polarized nucleon. For convenience we will consider two
projections of the nucleon spin vector, choosing the spin direction 
longitudinally and transversely to the nucleon momentum. In the nucleon 
rest frame one has
\begin{eqnarray}
S_L &=& (0, 0, 0, M)~, \nonumber\\
S_T &=& M (0, \cos\alpha, \sin\alpha,0)~.
\end{eqnarray}
The polarized part of the scattering cross section for longitudinal 
nucleon polarization, integrated over the azimuthal angle $\phi$, reads
\begin{eqnarray}
\frac{d^2\sigma(\lambda, \pm S_L)}{dxdy}
&=& \pm  
2 \pi S  \frac{\alpha^2}{Q^4}
\sum_i C_i \eta_i(Q^2) 
\nn\\
&\times& \left [
-2\lambda y \left( 2-y-\frac{2 x y M^2}{S} \right) xg_1^i +
8 \lambda \frac{y x^2 M^2}{S} g_2^i
 + \frac{4 x M^2}{S} \left ( 1 - y - \frac{x y M^2}{S} \right)
 g_3^i \right.
\nn\\
&-& 2 
\left (1 + \frac{2 x M^2}{S} \right)
\left (1 - y - \frac{x y M^2}{S} \right ) g_4^i
- 2 x y^2 \left (1 + \frac{2 x M^2}{S} \right) g_5^i \nn\\
&+&   \left.
4\lambda\frac{x y M^2}{S} g_6^i -2\left(1-y-\frac{x y M^2}{S}\right) 
g_9^i \right]~.
\label{scaL}
\ea
Correspondingly, for transversely polarized nucleons one obtains
\ba
\frac{d^3\sigma(\lambda, \pm S_T)} {dx dy d\phi} &=&
\pm
S \frac{\alpha^2}{Q^4} \sum_i
C_i \eta_i(Q^2)
\nn\\
&\times&\!\!\!
 2 \sqrt{\frac{M^2}{S}} \sqrt{x y \left [1 - y - \frac{x y M^2}{S}
\right]} \cos(\alpha-\phi) \left [
-2\lambda y x g_1^i -
4\lambda x  g_2^i \right.
\\
&-& \left.\!\!\!\frac{1}{y} \left (2 - y
- \frac{2 x y M^2}{S} \right) g_3^i
+ \frac{2}{y}  \left (1 - y - \frac{x y M^2}{S} \right) g_4^i
+  2 x y g_5^i -2\lambda g_6^i -g_9^i  \right]~.
\nn
\label{scaT}
\ea
Here $C^\gamma = 1$, $C^{\gamma Z} = g_V + \lambda g_A$, 
$C^Z = (g_V + \lambda g_A)^2$, and $C^{W^{\pm}} = (1 \pm \lambda)$.

As well--known, the contributions of the structure functions $g_2^i$ and
$g_3^i$ are suppressed by a factor of $M^2/S$ in the longitudinal spin 
asymmetries 
\ba
\Delta^L = d^2 \sigma(\lambda, S_L) - d^2 \sigma(\lambda, -S_L)~.
\ea
However, they contribute at the same strength as the other structure 
functions to the transverse spin asymmetries
\ba
\Delta^T = d^2 \sigma(\lambda, S_T) - d^2 \sigma(\lambda, -S_T)~,
\ea
which on their own behave $\propto M/\sqrt{S}$. The current 
non--conserving
structure functions $g_7^i$ and $g_8^i$ do not contribute to the cross 
sections when the lepton masses are neglected.
\section{The Forward Compton Amplitude}

\vspace{2mm}
\noindent
The forward Compton amplitude $T_{\mu\nu}$ is related to the hadronic 
tensor by
\begin{eqnarray}
W_{\mu\nu}^i = \frac{1}{2\pi} {\rm Im}~T_{\mu\nu}^i~,
\end{eqnarray}
where
\begin{eqnarray}
T^i_{\mu\nu} = i\int d^4x e^{iqx}\langle PS|
(T J_\mu^{i_1 \dagger}(x) J_\nu^{i_2}(0)|PS\rangle~.
\end{eqnarray}
The Compton amplitude can be represented in terms of the amplitudes 
$T_k^i$ and $A_k^i$, which are related to the unpolarized and polarized
structure functions, as
\begin{eqnarray}
T^i_{\mu\nu} =&& \left(-g_{\mu\nu}+\frac{q_\mu q_\nu}{q^2}\right)
T_1^i(q^2,\nu)
  + \frac{\hat P_\mu\hat P_\nu}{M^2} T_2^i(x,Q^2)
-i\varepsilon_{\mu\nu\lambda\sigma}\frac{q^\lambda P^\sigma}{2 M^2} 
T_3^i(q^2,\nu)
\nonumber \\
&& + \frac{q_\mu q_\nu}{M^2} T_4^i(q^2,\nu)
+ \frac{(p_\mu q_\nu+p_\nu q_\mu)}{2 M^2} T_5^i(q^2,\nu) 
\nonumber \\
&& + i\varepsilon_{\mu\nu\lambda\sigma}\frac{q^\lambda S^\sigma}{M^2} 
A_1^i(q^2,\nu) 
+i\varepsilon_{\mu\nu\lambda\sigma}
\frac{q^\lambda (P\cdot q S^\sigma-S\cdot q P^\sigma)}{M^4} 
A_2^i(q^2,\nu) 
\nonumber \\
&& + \left[\frac{\hat P_\mu\hat S_\nu+\hat S_\mu\hat P_\mu}{2} -
S\cdot q\frac{\hat P_\mu\hat P_\nu}{P\cdot q} \right] 
\frac{A_3^i(q^2,\nu)}
{M^2}\nonumber \\
&& + S \cdot q \frac{\hat P_\mu\hat P_\nu}{M^4} A_4^i(q^2,\nu) +
\left(-g_{\mu\nu}+\frac{q_\mu q_\nu}{q^2}\right) \frac{(S\cdot q)}{M^2} 
A_5^i(q^2,\nu),
\nonumber\\
&& + i\varepsilon_{\mu\nu\lambda\sigma} \frac{P_\sigma S_\lambda}{M^2} 
A_6^i(q^2,\nu) +
S\cdot q\frac{q_\mu q_\nu}{M^4} A_7^i(q^2,\nu) \nonumber \\
&&+\frac{(P_\mu q_\nu+q_\mu P_\nu) S\cdot q}{2 M^4} A_8^i(q^2,\nu)
+\frac{S_\mu q_\nu+S_\nu q_\mu}{2 M^2} A_9^i(q^2,\nu)~.
\end{eqnarray}
The structure functions $F_i(x,Q^2)$ and $g_i(x,Q^2)$ are obtained
from the amplitudes $T_i(q^2,\nu)$ and $A_i(q^2,\nu)$  by
\begin{eqnarray}
F_{1}(x,Q^2) & = & \frac{1}{2\pi} {\rm Im}~T_{1}(q^2,\nu)~,
\nonumber\\
F_{2,3,4,5}(x,Q^2) & = & \frac{1}{2\pi}\frac{\nu}{M^2}
{\rm Im}~T_{2,3,4,5}(q^2,\nu)~,
\label{relSIn}
\end{eqnarray}
for the unpolarized structure functions, and
\begin{eqnarray}
g_{1,3,5,6,9}(x,Q^2) & = & \frac{1}{2\pi}\frac{\nu}{M^2}
{\rm Im} A_{1,3,5,6,9}(q^2,\nu)~, \nonumber\\
g_{2,4,7,8}(x,Q^2) & = & \frac{1}{2\pi} \frac{\nu^2}{M^4}
{\rm Im}~A_{2,4,7,8}(q^2,\nu)~,
\label{relSD}
\end{eqnarray}
for the polarized structure functions. In the following we will consider
the polarized part of $T_{\mu\nu}^i$ only.

For neutral current interactions the current operators obey
\be
{J_\mu^{\gamma, Z}}^\dagger = J_\mu^{\gamma, Z}~.
\label{neut}
\ee
Therefore, the crossing relation for the amplitude for $q \rightarrow -q,
P \rightarrow P$ yields
\be
\label{eqCR1}
T_{\mu\nu}^{i}(q^2, -\nu) = T_{\nu\mu}^{i}(q^2, \nu)~.
\ee
The corresponding relations
for the amplitudes $A_i^{\rm NC}(q^2, \nu)$ are
\begin{eqnarray}
A_{1,3,8,9}^{\rm NC}(q^2, -\nu) &=& ~A_{1,3,8,9}^{\rm NC}(q^2, \nu)~,
 \nonumber\\
A_{2,4,5,6,7}^{\rm NC}(q^2, -\nu) &=& -A_{2,4,5,6,7}^{\rm NC}(q^2, \nu)~.
\end{eqnarray}
Furthermore, the amplitudes obey the following forward dispersion 
relations
\begin{eqnarray}
A_{1,3,8,9}^{\rm NC}(q^2, \nu) &=& 
\frac{2}{\pi} \int_{Q^2/2}^{\infty} d\nu' \frac{\nu'} {\nu'^2 - \nu^2}
 {\rm Im}~A_{1,3,8,9}^{\rm NC}(q^2, \nu^\prime)~, \nn\\
A_{2,4,5,6,7}^{\rm NC}(q^2, \nu) &=& \frac{2}{\pi}
\int_{Q^2/2}^{\infty} d \nu' \frac{\nu}
{\nu'^2 - \nu^2} {\rm Im}~A_{2,4,5,6,7}^{\rm NC}(q^2, \nu')~.
\label{ndr}
\end{eqnarray}

For the charged current interactions it is more suitable to study the 
linear combination of amplitudes
\ba
\label{eqCOM}
T_{\mu\nu}^{\pm}(q^2, \nu) = T_{\mu\nu}^{W^-}(q^2, \nu)
                        \pm T_{\mu\nu}^{W^+}(q^2, \nu)~.
\ea
Due to the transformation
\be
{J_\mu^{W^\pm}}^\dagger= J_\mu^{W^\mp}~,
\label{char}
\ee
the crossing relations read
\be
T^{\pm}(q^2, -\nu) = \pm T^{\pm}(q^2, \nu)~.
\ee
Correspondingly, one obtains for the combination of the amplitudes
\be
A_i^{\pm}={A_i}^{W^-}\pm {A_i}^{W^+}
\ee
the relations
\ba
A^\pm_{1,3,8,9}(q^2,-\nu) &=&
\pm A^\pm_{1,3,8,9}(q^2,\nu)~,{\ }{\ }\nn \\
A^\pm_{2,4,5,6,7}(q^2,-\nu) &=&
\mp A^\pm_{2,4,5,6,7}(q^2,\nu)~.
\label{cr}
\ea
The respective dispersion relations for the amplitudes $A_i^+(q^2,\nu)$ 
and $A_i^-(q^2,\nu)$ are
\begin{eqnarray}
A_{1,3,8,9}^+(q^2, \nu) &=& \frac{2}{\pi} \int_{Q^2/2}^{\infty} d
\nu' \frac{\nu'} {\nu'^2 - \nu^2} {\rm Im} A_{1,3,8,9}^+(q^2,
\nu^\prime)~,\nn \\
A_{2,4,5,6,7}^+(q^2, \nu)&=& \frac{2}{\pi}
\int_{Q^2/2}^{\infty} d \nu' \frac{\nu}
{\nu'^2 - \nu^2} {\rm Im} A_{2,4,5}^+(q^2, \nu')~,\nn\\
A_{1,3,8,9}^-(q^2, \nu) &=& \frac{2}{\pi} \int_{Q^2/2}^{\infty} d
\nu' \frac{\nu} {\nu'^2 - \nu^2} {\rm Im}
 A_{1,3,8,9}^-(q^2, \nu^\prime)~, \nn \\
A_{2,4,5,6,7}^-(q^2, \nu) &=& \frac{2}{\pi} \int_{Q^2/2}^{\infty} d
\nu' \frac{\nu'} {\nu'^2 - \nu^2} {\rm Im}
 A_{2,4,5,6,7}^-(q^2, \nu^\prime)~.
\label{cdr2}
\end{eqnarray}
In the case of charged current interactions we introduce the structure
function combinations
\be
g_i^{\pm}(x,Q^2) = g_i^{W^-}(x,Q^2) \pm g_i^{W^+}(x,Q^2)~.
\label{eqGIPM}
\ee
The integral representations of the amplitudes $A_i^{\rm NC}$ and
$A_i^{\pm}$ can be finally expressed by the moments of the corresponding
structure functions as
\ba
A_{1,3,9}^{NC,+}(q^2,\nu)&=&\frac{4M^2}{\nu}
\sum_{n=0,2...}\frac{1}{x^{n+1}}
\int_0^1dyy^ng_{1,3,9}^{NC,+}(y,Q^2)~,\nn\\
A_{2,4,7}^{NC,+}(q^2,\nu)&=&\frac{4M^4}{\nu^2}
\sum_{n=2,4...}\frac{1}{x^{n+1}}
\int_0^1dyy^ng_{2,4,7}^{NC,+}(y,Q^2)~,\nn\\
A_{5,6}^{NC,+}(q^2,\nu)&=&\frac{4M^2}{\nu}
\sum_{n=1,3...}\frac{1}{x^{n+1}}\int_0^1dyy^ng_{5,6}^{NC,+}(y,Q^2)~,
\nn\\
A_{8}^{NC,+}(q^2,\nu)&=&\frac{4M^4}{\nu^2}
\sum_{n=1,3...}\frac{1}{x^{n+1}}\int_0^1dyy^ng_{8}^{NC,+}(y,Q^2)~,
\label{tayNC}
\ea
and
\ba
A_{1,3,9}^{-}(q^2,\nu)&=&\frac{4M^2}{\nu}
\sum_{n=1,3...}\frac{1}{x^{n+1}}
\int_0^1dyy^ng_{1,3,9}^{-}(y,Q^2)~,\nn\\
A_{2,4,7}^{-}(q^2,\nu)&=&\frac{4M^4}{\nu^2}
\sum_{n=1,3...}\frac{1}{x^{n+1}}
\int_0^1dyy^ng_{2,4}^{-}(y,Q^2)~,
\nn\\
A_{5,6}^{-}(q^2,\nu)&=&\frac{4M^2}{\nu}
\sum_{n=0,2...}\frac{1}{x^{n+1}}
\int_0^1dyy^ng_{5,6}^{-}(y,Q^2)~,
\nn\\
A_{8}^{-}(q^2,\nu)&=&\frac{4M^4}{\nu^2}
\sum_{n=2,4...}\frac{1}{x^{n+1}}
\int_0^1dyy^ng_{8}^{-}(y,Q^2)~,
\label{tay-}
\ea
performing a Taylor expansion of Eqs.~(\ref{ndr}~, \ref{cdr2}) and using
Eqs.~(\ref{relSD}~, \ref{eqGIPM}), where $x = Q^2/(2 \nu)$ and
the integration variable $y = Q^2/(2 \nu')$~.

\section{Operator product expansion}

\vspace{2mm}
\noindent
The operator product expansion is one of the most general formalisms to
analyze the properties of the structure functions in deep-inelastic 
scattering. We apply it to the $T$-product of two electroweak currents,
\begin{eqnarray}
\widehat T_{\mu\nu}^i = T(J_\mu^{i_1\dagger}(x)J_{\nu}^{i_2}(0))~.
\end{eqnarray}
Near the light cone one obtains for neutral currents
\begin{eqnarray}
\widehat T^{NC}_{\mu\nu}& =& \bar q(x)\gamma_\mu(g_{V_1}+g_{A_1}\gamma_5)
S(\,x) \gamma_\nu(g_{V_2}+g_{A_a}\gamma_5)P^+q(0)
\nonumber \\
& +& \bar q(0)\gamma_\nu(g_{V_2}+g_{A_2}\gamma_5)S(\,x)
\gamma_\mu(g_{V_1}+g_{A_1}\gamma_5)P^+q(x)~,
\label{TNC1}
\end{eqnarray}
and for the charged current combinations 
\begin{eqnarray}
\widehat T^{\pm}_{\mu\nu} = \widehat 
T^{W_-}_{\mu\nu}\pm \widehat T^{W^+}_{\mu\nu}~,
\end{eqnarray}
where
\begin{eqnarray}
\widehat T^{\pm}_{\mu\nu}& =& 
\bar q(x)\gamma_\mu(g_{V_1}+g_{A_1}\gamma_5)
S(\,x)
\gamma_\nu(g_{V_2}+g_{A_a}\gamma_5)P^{\pm}q(0)
\nonumber \\
& \pm& \bar q(0)\gamma_\nu(g_{V_2}+g_{A_2}\gamma_5)S(\,x)
\gamma_\mu(g_{V_1}+g_{A_1}\gamma_5)P^{\pm}q(x)~.
\label{TPM1}
\end{eqnarray}
Here we used the projectors
\be
P^+= {\bf 1} ,~~~~~~~~~~~~~~~~~~~~~~~P^-= {\bf \tau_3} =
\left ( \begin{array}{rr} 1 & 0 \\
0 & -1 \end{array} \right)~.
\ee 
The Fourier transforms of the expressions (\ref{TNC1}) and (\ref{TPM1}) 
have the following form
\begin{eqnarray}
i\int d^4x e^{iqx}  \widehat
T_{\mu\nu}^{NC} &=&
 \nonumber\\
&-& \int\frac{d^4k}{(2\pi)^4}
\overline U(k)\gamma_\mu(g_{V_1}+g_{A_1}\gamma_5)
\frac{\hat k+\hat q+m_q}{(k+q)^2-m_q^2}
\gamma_\nu(g_{V_2}+g_{A_2}\gamma_5)P^+ U(k)\nonumber\\
&-& \int\frac{d^4k}{(2\pi)^4}
\overline U(k)\gamma_\mu(g_{V_1}+g_{A_1}\gamma_5)
\frac{\hat k-\hat q+m_q}{(k-q)^2-m_q^2}
\gamma_\nu(g_{V_2}+g_{A_2}\gamma_5)P^+ U(k),
\label{FourTNC}
\end{eqnarray}
and
\begin{eqnarray}
i\int d^4x e^{iqx}  \widehat
T_{\mu\nu}^{\pm}&=& 
\nonumber\\
&-& \int\frac{d^4k}{(2\pi)^4}
\overline U(k)\gamma_\mu(g_{V_1}+g_{A_1}\gamma_5)
\frac{\hat k+\hat q+m_q}{(k+q)^2-m_q^2}
\gamma_\nu(g_{V_2}+g_{A_2}\gamma_5)P^{\pm} U(k)\nonumber\\
&\mp& \int\frac{d^4k}{(2\pi)^4}
\overline U(k)\gamma_\mu(g_{V_1}+g_{A_1}\gamma_5)
\frac{\hat k-\hat q+m_q}{(k-q)^2-m_q^2}
\gamma_\nu(g_{V_2}+g_{A_2}\gamma_5)P^{\pm} U(k).
\label{FourTPM}
\end{eqnarray}
Here $m_q$ denotes the mass of the outgoing quark in the scattering
process and
\begin{eqnarray}
U(k)=\int{ \frac{d^4k}{(2\pi)^4}~e^{-ikx}\psi(x)}~.
\end{eqnarray}
In the case of forward scattering the Fourier transform of the
current product depends only on one variable.
 
So far our discussion was quite general. In the following we will
disregard the quark mass effects. As a consequence current conservation
is also obtained for weak currents and only the structure functions
$\left. F_i\right|_{i=1}^3$ and $\left. g_i \right|_{i=1}^5$ contribute.
To isolate the spin dependent part in Eqs.~(\ref{FourTNC}) and 
(\ref{FourTPM}) we use the identities
\begin{eqnarray}
\gamma_\mu\hat a\gamma_\nu &=& a^\alpha[S_{\mu\alpha\nu\beta}
\gamma_\beta-
i\varepsilon_{\mu\alpha\nu\beta}\gamma_\beta\gamma_5]~,
\nonumber\\ 
S_{\mu\alpha\nu\beta} & = & g_{\mu\alpha}g_{\mu\beta}
+g_{\mu\beta}g_{\nu\alpha}-g_{\mu\nu}g_{\alpha\beta}~.
\end{eqnarray}
For the spin dependent part of $\widehat T_{\mu\nu}^i$ one obtains
\begin{eqnarray}
\widehat T^+_{\mu\nu} &=& -i(g_{V_1}g_{V_2}+g_{A_1}g_{A_2})\epa 
q^\alpha u_+^\beta+
(g_{V_1}g_{A_2}+g_{A_1}g_{V_2})\Sa [q^\alpha u_-^\beta 
+u^{\alpha\beta}]~,\\
\widehat T^-_{\mu\nu} &=& -i(g_{V_1}g_{V_2}+g_{A_1}g_{A_2})\epa 
q^\alpha v_-^\beta+
(g_{V_1}g_{A_2}+g_{A_1}g_{V_2})\Sa [q^\alpha v_+^\beta 
+v^{\alpha\beta}]~,
\end{eqnarray}
with
\begin{eqnarray}
u^\beta_{\pm} &=& -
\int\frac{d^4k}{(2\pi)^4}\overline U(k)
\frac{\gamma_\beta \gamma_5}{(k+q)^2}P^+U(k)~
\mp ~(q\leftrightarrow-q)~,\\
u^{\alpha\beta} & =& -
\int\frac{d^4k}{(2\pi)^4}\overline U(k)
\frac{k_\alpha\gamma_\beta \gamma_5}{(k+q)^2}P^+U(k)~
- ~(q\leftrightarrow-q)~,\\
v^\beta_{\pm} &=& -
\int\frac{d^4k}{(2\pi)^4}\overline U(k)
\frac{\gamma_\beta \gamma_5}{(k+q)^2}P^-U(k)~
\mp ~(q\leftrightarrow-q)~,\\
v^{\alpha\beta} & =& -
\int\frac{d^4k}{(2\pi)^4}\overline U(k)
\frac{k_\alpha\gamma_\beta \gamma_5}{(k+q)^2}P^-U(k)~
- ~(q\leftrightarrow-q)
\end{eqnarray}
in the massless quark limit $m_q \rightarrow 0$. Expanding the 
denominators $(k+q)^2$ and $(k-q)^2$ in the above relations into powers of 
the ratio $2k.q/Q^2$ the corresponding operators can be expressed 
in terms of the two operators $\Theta^{\pm\beta\{\mu_1,\cdots\mu_n\}}$~:
\begin{eqnarray}
u^\beta_{\pm} &=&
\sum_{n~even,odd}{\biggl(\frac{2}{Q^2}\biggr)^{n+1} q_{\mu_1}\ldots 
q_{\mu_n}
\Theta^{\pm\beta\{\mu_1\cdots\mu_n\}}}~,\label{ub}\\ 
u^{\alpha\beta} &=&
\sum_{n~even}{\biggl(\frac{2}{Q^2}\biggr)^{n+1} q_{\mu_1}\ldots 
q_{\mu_n}
\Theta^{+\beta\{\alpha\mu_1\cdots\mu_n\}}}~,\label{uab}\\ 
v^\beta_{\pm} &=&
\sum_{n~even,odd}{\biggl(\frac{2}{Q^2}\biggr)^{n+1} q_{\mu_1}\ldots 
q_{\mu_n}
\Theta^{\mp\beta\{\mu_1\cdots\mu_n\}}}~,\label{vb}\\ 
v^{\alpha\beta} &=&
\sum_{n~odd}{\biggl(\frac{2}{Q^2}\biggr)^{n+1} q_{\mu_1}\ldots 
q_{\mu_n}
\Theta^{-\beta\{\alpha\mu_1\cdots\mu_n\}}}~,\label{vab} 
\end{eqnarray}
where
\begin{eqnarray}
\Theta^{\pm\beta\{\mu_1\cdots\mu_n\}} = 
\int{\frac{d^4k}{(2\pi)^4} \overline U(k)\gamma_\beta\gamma_5
k_{\mu_1}\ldots k_{\mu_n}P^{\pm}U(k)}~.
\end{eqnarray}
Inserting the expressions for the quark operators (\ref{ub}--\ref{vab})
into the spin--dependent part of the $T$-product, $\widehat T^i_{spin}$, 
one obtains
\begin{eqnarray}
\widehat T^+_{\mu\nu,spin}
&=& -i(g_{V_1}g_{V_2}+g_{A_1}g_{A_2})
\varepsilon_{\mu\alpha\nu\beta} q^\alpha 
\sum_{n~even} q^{\mu_1} \cdots q^{\mu_n} 
\left(\frac{2}{Q^2}\right)^{n+1}
\Theta^{+\beta\{\mu_1\cdots\mu_n\}}
\nonumber\\
&+& (g_{V_1}g_{A_2}+g_{V_2}g_{A_1})
\left[-g_{\mu\nu}
\sum_{n~even} q^{\mu_1} \cdots q^{\mu_n} 
\left(\frac{2}{Q^2}\right)^{n} 
\Theta^{+\mu_1\{\mu_2\cdots\mu_n\}} \right.
\nonumber\\
&+&\sum_{n~odd} q^{\mu_1} \cdots q^{\mu_n} 
\left(\frac{2}{Q^2}\right)^{n+1} +
\left(q_\mu \Theta^{+\nu\{\mu_1\cdots\mu_n\}}
+q_\nu\Theta^{+\mu\{\mu_1\cdots\mu_n\}} \right)
\nonumber \\
&+& \left.
\sum_{n~even} q^{\mu_1} \cdots q^{\mu_n} \left(\frac{2}{Q^2}
\right)^{n+1}\left(\Theta^{+\mu\{\nu\mu_1\cdots\mu_n\}}
+\Theta^{+\nu\{\mu\mu_1\cdots\mu_n\}}
\right)\right]~,
\label{TCompP}
\end{eqnarray}
and
\begin{eqnarray}
\widehat T^-_{\mu\nu,spin}&=& -i(g_{V_1}g_{V_2}+g_{A_1}g_{A_2})
\varepsilon_{\mu\alpha\nu\beta}q^\alpha 
\sum_{n~odd}{q^{\mu_1}\cdots q^{\mu_n} \left(\frac{2}{Q^2}\right)^{n+1}
\Theta^{-\beta\{\mu_1\cdots\mu_n\}}}\nonumber\\
&+& (g_{V_1}g_{A_2}+g_{V_2}g_{A_1})\left[-g_{\mu\nu}
\sum_{n~odd}{q^{\mu_1}\cdots q^{\mu_n} \left(\frac{2}{Q^2}\right)^{n}
\Theta^{-\mu_1\{\mu_2\cdots\mu_n\}}}\right.
\nonumber\\
&+& \sum_{n~even}{q^{\mu_1}\cdots q^{\mu_n} \left(\frac{2}{Q^2}
\right)^{n+1}  \left(q_\mu\Theta^{-\nu\{\mu_1\cdots\mu_n\}} 
+q_\nu\Theta^{-\mu\{\mu_1\cdots\mu_n\}}\right)}
\nonumber \\
&+& \left.
\sum_{n~odd} q^{\mu_1}\cdots q^{\mu_n} \left(\frac{2}{Q^2}
\right)^{n+1} 
\left(\Theta^{-\mu\{\nu\mu_1\cdots\mu_n\}} 
+\Theta^{-\nu\{\mu\mu_1\cdots\mu_n\}}\right)\right]~.
\label{TCompM}
\end{eqnarray}
The operators $\Theta^{\pm\beta\{\mu_1\cdots\mu_n\}}$ can be decomposed 
into a symmetric part, $\Theta_S$, and a remainder, $\Theta_R$, 
respectively
\begin{eqnarray}
\Theta^{\pm\beta\{\mu_1\cdots\mu_n\}}
= \Theta_S^{\pm\beta\{\mu_1\cdots\mu_n\}}+
\Theta_R^{\pm\beta\{\mu_1]\cdots\mu_n\}}~,
\end{eqnarray}
where
\begin{eqnarray}
 \Theta_S^{\pm\beta\{\mu_1\cdots\mu_n\}} &=&
  \frac{1}{n+1} \left[\Theta^{\pm\beta\{\mu_1\cdots\mu_n\}}+
                      \Theta^{\pm\mu_1\{\beta\cdots\mu_n\}}+\ldots+
                      \Theta^{\pm\mu_n\{\mu_1\cdots\beta\}}\right]~,\\
 \Theta_R^{\pm\beta\{\mu_1\cdots\mu_n\}} &=&
  \frac{1}{n+1} \left[\Theta^{\pm\beta\{\mu_1\cdots\mu_n\}}
                     -\Theta^{\pm\mu_1\{\beta\cdots\mu_n\}}+
                      \Theta^{\pm\beta\{\mu_1\mu_2\cdots\mu_n\}}
                     -\Theta^{\pm\mu_2\{\mu_1\beta\cdots\mu_n\}}+
\ldots\right].
\end{eqnarray}
In the massless quark limit both operators $\Theta_S$ and $\Theta_R$
are traceless. To find the target mass dependence of the structure
functions we have to construct the traceless nucleon matrix elements of 
these operators~\cite{GP}.

The rank-$n$ symmetric, traceless tensor constructed from $n$ nucleon
momenta $P_{\mu_i}$~~($P^2 = M^2$) has the form
\begin{eqnarray}
\Pi^{\mu_1\cdots\mu_n}(P) &=& \sum_{j=0}^{[n/2]}
{\frac{(-1)^j}{2^j}\frac{(n-j)!}{n!}\underbrace{g\cdots g}_{j}~
\underbrace{P\cdots P}_{n - 2j}  M^{2j}}~.
\end{eqnarray}
Here, $[n/2]=n/2$ for even $n$ and $[n/2]=(n-1)/2$ for odd $n$. The sum 
contains $j$ metric tensors $g_{\mu_i\nu_k}$ with indices chosen in the
set $\{\mu_1,\ldots,\mu_n\}$ in all possible combinations. The remaining 
$n - 2j$ indices are carried by the momenta $P_{\mu_l}$. The sum contains
$n!/[2^jj!(n-2j)!]$ terms.

For the representation of the $\Theta$-operators traceless and symmetric 
tensors have to be constructed from one nucleon spin vector $S$ and $n$
momentum vectors $P$. These tensors have the form~\cite{GS}
\begin{eqnarray}
M_1^{\mu_1\cdots\mu_n}(P,S) &\equiv&
\left\{S^{\mu_1}P^{\mu_2}\ldots P^{\mu_n}\right\}\nonumber\\ 
&=&
\frac{1}{n}\sum_{i}{S^{\mu_i} \Pi^{\mu_1\cdots[\mu_i]\cdots\mu_n}}(P)
-
\frac{1}{n^2}
\sum_{i<j}{g_{\mu_i\mu_j}S_\alpha
\Pi^{\mu_1\cdots[\mu_i]\cdots[\mu_j]\cdots\mu_n}(P)}~,
\end{eqnarray}
where the indices in brackets $[\mu_i]$ indicate that the corresponding 
superscript is to be removed from the sequence. The coefficient of the
second term in the above expression is chosen such that the trace vanishes 
identically. A more compact expression for these tensors was obtained in 
Ref.~\cite{PR},
\begin{eqnarray}
M_1^{\mu_1\cdots\mu_n}(P,S) =
\left\{S^{\mu_1}P^{\mu_2}\ldots P^{\mu_n}\right\} & =&
 \frac{1}{n}\sum_{j=0}^{[(n-1)/2]}
{\frac{(-1)^j}{2^j}\frac{(n-j)!}{n!}\underbrace{g\cdots g}_{j}~
\underbrace{[SP\cdots P]_S}_{n - 2j}
M^{2j}}~.
\end{eqnarray}
Again the  sum contains $j$ times the metric tensors $g_{\mu_i\mu_k}$.
The remaining $n-2j$  indices are symmetrized in the product 
$[SP\ldots P]_S$.
\section{The Nucleon Matrix Elements} 

\vspace{2mm}
\noindent
We calculate now the nucleon matrix element of the operators derived in
the preceding section. The expectation value of the symmetric operators
$\Theta_S$ is given by
\begin{eqnarray}
\langle PS|\Theta_S^{\pm\beta\{\mu_1\cdots\mu_n\}}|PS\rangle & = & 
a_n^{\pm}
 M_1^{\beta\mu_1\cdots\mu_n}~.
\end{eqnarray}
Correspondingly, for the operators $\Theta_R$ one obtains, cf. 
also~\cite{PR}, 
\begin{eqnarray}
\langle PS|\Theta_R^{\pm\beta\{\mu_1\cdots\mu_n\}}|PS\rangle  = 
\frac{d_n^{\pm}}{n+1}\Big[&&
M_2^{\beta\{\mu_1\mu_2\cdots\mu_n\}}-
 M_2^{\mu_1\{\beta\mu_2\cdots\mu_n\}}+
\ldots \nonumber\\
&& M_2^{\beta\{\mu_1\mu_2\cdots\mu_n\}}
- M_2^{\mu_n\{\mu_1\mu_2\cdots\beta\}}
~~~~~\Big]~,
\label{Theta_R}
\end{eqnarray}
where 
\begin{eqnarray}
M_2^{\beta\mu_1\cdots\mu_n}(P,S) = \frac{n+2}{n+1}S^\beta\Pi^{\mu_1
\cdots\mu_n}(P) + \frac{n}{n+1}P^\beta M_1^{\mu_1\cdots\mu_n}(P,S)~.
\end{eqnarray}
In the massless nucleon limit the known results, cf. Ref.~~\cite{BK}~,
are reproduced,
\ba
\label{MS}
\langle PS | \Theta_S^{\pm\beta\{\mu_1 ... \mu_n\}} |PS\rangle &=&
\frac{a_n^\pm}{n+1} \left [ S^{\beta} P^{\mu_1} P^{\mu_2} ... P^{\mu_n}
                      + S^{\mu1} P^{\beta} P^{\mu_2} ... P^{\mu_n}
                      + ...  \right ]~, \\
\label{MR}
\langle PS | \Theta_R^{\pm\beta\{\mu_1] ... \mu_n\}} |PS\rangle &=&
\frac{d_n^\pm}{n+1} 
\left [ \left( S^{\beta} P^{\mu_1} - S^{\mu_1} P^{\beta}
\right ) P^{\mu_2} ... P^{\mu_n}  \right.
\nonumber\\
& &~~~~~~+
\left( S^{\beta} P^{\mu_2} - S^{\mu_2} P^{\beta}
\right ) P^{\mu_1} P^{\mu_3} ... P^{\mu_n}  \nonumber\\
& &~~~~~~+ \left.
\ldots + \left( S^{\beta} P^{\mu_n} - S^{\mu_n} P^{\beta}
\right ) P^{\mu_1} P^{\mu_3} ... P^{\mu_{n-1}}  \right ]~.
\ea
$a_n^{\pm}$ and $d_n^{\pm}$ denote the twist--2 and twist--3 operator
matrix elements in the case of neutral current ($+,NC$) and charged 
current interactions ($\pm$) of moment index $n$. These are 
non--perturbative quantities and are independent of the nucleon mass. All
information about the target mass corrections is contained in the tensors
$M_1^{\beta\mu_1 ... \mu_n}$ and $M_2^{\beta\mu_1 ... \mu_n}$~etc.

Let us consider first the nucleon matrix element of the first term of
the operator $\widehat T^{+,1}_{\mu\nu,\,spin}$, Eq.~(\ref{TCompP}), 
\begin{eqnarray}
\widehat T^{+,1}_{\mu\nu,spin}&=& -i(g_{V_1}g_{V_2}+g_{A_1}g_{A_2})
\varepsilon_{\mu\alpha\nu\beta}q^\alpha 
\sum_{n~even}{q^{\mu_1}\cdots q^{\mu_n} \Bigl(\frac{2}{Q^2}\Bigr)^{n+1}
\Theta^{+\beta\{\mu_1\cdots\mu_n\}}}.
\label{T1forWW}
\end{eqnarray}
to derive the structure of the target mass correction for forward Compton 
scattering. We, furthermore, consider only the symmetric part of the 
operator $\Theta$.
\begin{eqnarray}
\langle PS|\widehat T^{+,1}_{\mu\nu,spin}|PS\rangle &=& 
-i(g_{V_1}g_{V_2}+g_{A_1}g_{A_2})\epa q^{\alpha}\nonumber\\
&\times& \Bigg\{\sum_{n~even}{\sum_{j=0}^{[n/2]}{\frac{1}{x^{(n-2j+1)}}
\left(\frac{M^2}{Q^2}\right)^j\frac{S^\beta}{\nu}\frac{(n-j+1)!}{j!
(n-2j)!}
\frac{a_n^+}{(n+1)^2}}}\nonumber \\
&&+\sum_{n~even}{\sum_{j=0}^{[n/2]}{\frac{1}{x^{(n-2j+1)}}
\left(\frac{M^2}{Q^2}\right)^j\frac{P^\beta (S\cdot q)}{\nu^2}\frac{
(n-j+1)!}
{j!(n-2j-1)!}
\frac{a_n^+}{(n+1)^2}}}\Bigg\}~.
\end{eqnarray}
Shifting the summation index from $n$ to $n=n+2j$ both sums become
independent and one obtains
\begin{eqnarray}
\langle PS|\widehat T^{+,1}_{\mu\nu,spin}|PS\rangle &=& 
-i(g_{V_1}g_{V-2}+g_{A_1}g_{A_2})\epa q^{\alpha}\nonumber\\
&\times&
\Bigg\{\sum_{n~even}{\sum_{j=0}^{\infty}{\frac{1}{x^{(n+1)}}
\left(\frac{M^2}{Q^2}\right)^j\frac{S^\beta}{\nu}\frac{(n+j+1)!}{j!\,n!}
\frac{a_{n+2j}^+}{(n+2j+1)^2}}}\nonumber \\
&&+\sum_{n~even}{\sum_{j=0}^{\infty}{\frac{1}{x^{(n+1)}}
\left(\frac{M^2}{Q^2}\right)^j\frac{P^\beta (S\cdot q)}{\nu^2}\frac{
(n+j+1)!}
{j!(n-1)!}
\frac{a_{n+2j}^+}{(n+2j+1)^2}}}\Bigg\}~.
\end{eqnarray}
Carrying out the same steps also for the other parts of the expressions
(\ref{TCompP}) and (\ref{TCompM}) and rearranging the tensor structures 
according to the structure functions one obtains for the twist--$2$ part 
of the forward Compton amplitude the relations
\begin{eqnarray}
\langle PS|\widehat T_{\mu\nu, spin}^{NC,+,~\tau=2}|PS\rangle &=& 
-i (g_{V_1}g_{V_2}+g_{A_1}g_{A_2})
\bigg\{
\varepsilon_{\mu\alpha\nu\beta}\frac{q^\alpha S^\beta}{\nu}
\sum_{n~even}{\frac{1}{x^{n+1}}
(n+1)\sum_{j=0}^{\infty}
{\cal C}(n,j)~a^+_{n+2j}}
\nonumber\\
&-&\varepsilon_{\mu\alpha\nu\beta}
\frac{q^\alpha P^\beta(S\cdot q)-q^\alpha S^\beta(P\cdot q)}{\nu^2}
\sum_{n~even}{\frac{1}{x^{n+1}}
n\sum_{j=0}^{\infty}{
{\cal C}(n,j)~a^+_{n+2j}}}
\bigg\}
\nonumber\\
& +&(g_{V_1}g_{A_2}+g_{V_2}g_{A_1})
 \left\{\left(-g_{\mu\nu}+\frac{q_\mu q_\nu}{q^2}\right)
 \frac{S\cdot q}{\nu}\right.
 \nonumber\\
&&\sum_{n~odd}\frac{1}{x^{n+1}}
\sum_{j=0}^{\infty}
{\cal C}(n,j)\frac{a^+_{n+2j}}{n+2j} 
[(n+2j+1)(n+2j)+2j]\nonumber\\
&+&\left[\frac{\hat S^\mu\hat P^\nu+\hat P^\mu\hat S^\nu}{2}-
\frac{\hat P^\mu\hat P^\nu}{\nu}(S\cdot q)\right]
\frac{4}{\nu}\sum_{n~odd}{\frac{1}{x^n}
n\sum_{j=0}^{\infty}
{\cal C}(n,j)\frac{a^+_{n+2j}}{n+2j}}\nonumber\\
&+&\frac{\hat P^\mu\hat P^\nu}{\nu}(S\cdot q)
\frac{2}{\nu}\sum_{n~odd}\frac{1}{x^n}
n(n+1)\sum_{j=0}^{\infty}
{\cal C}(n,j)\frac{a^+_{n+2j}}{n+2j}\bigg\}~,
\label{TCompNCt2}
\end{eqnarray}
and
\begin{eqnarray}
\langle PS|\widehat T_{\mu\nu, spin}^{-,~\tau=2}|PS\rangle &=& 
-i (g_{V_1}g_{V_2}+g_{A_1}g_{A_2})
\bigg\{\varepsilon_{\mu\alpha\nu\beta}\frac{q^\alpha S^\beta}{\nu}
\sum_{n~odd}{\frac{1}{x^{n+1}}
(n+1)\sum_{j=0}^{\infty}
{\cal C}(n,j)~a^-_{n+2j}}\nonumber\\
&-&\varepsilon_{\mu\alpha\nu\beta}
\frac{q^\alpha P^\beta(S\cdot q)-q^\alpha S^\beta(P\cdot q)}{\nu^2}
\sum_{n~odd}{\frac{1}{x^{n+1}}
n\sum_{j=0}^{\infty}{
{\cal C}(n,j)~a^-_{n+2j}}}
\bigg\}
\nonumber\\
&&~~~~~~~~~~~ +(g_{V_1}g_{A_2}+g_{V_2}g_{A_1})
 \bigg\{(-g_{\mu\nu}+\frac{q_\mu q_\nu}{q^2})\frac{S\cdot q}{\nu} 
 \nonumber\\
&&\sum_{n~even}\frac{1}{x^{n+1}}
\sum_{j=0}^{\infty}
{\cal C}(n,j)\frac{a^-_{n+2j}}{n+2j} 
[(n+2j+1)(n+2j)+2j]
\nonumber\\
&+&\left[\frac{\hat S^\mu\hat P^\nu+\hat P^\mu\hat S^\nu}{2}-
\frac{\hat P^\mu\hat P^\nu}{\nu}(S\cdot q)\right]
\frac{4}{\nu}\sum_{n~even}{\frac{1}{x^n}
n\sum_{j=0}^{\infty}
{\cal C}(n,j)\frac{a^-_{n+2j}}{n+2j}}\nonumber\\
&+&\frac{\hat P^\mu\hat P^\nu}{\nu}(S\cdot q)
\frac{2}{\nu}\sum_{n~even}\frac{1}{x^n}
n(n+1)\sum_{j=0}^{\infty}
{\cal C}(n,j)\frac{a^-_{n+2j}}{n+2j}\bigg\}~.
\label{TComp-t2}
\end{eqnarray}

Correspondingly, the twist--3 contributions read
\begin{eqnarray}
\langle PS|\widehat T_{\mu\nu, spin}^{NC,+,~\tau=3}|PS\rangle 
&=& 
\nonumber\\ & &
\hspace*{-1cm}
-i(g_{V_1}g_{V_2}+g_{A_1}
g_{A_2})
\bigg\{\varepsilon_{\mu\alpha\nu\beta}\frac{q^\alpha S^\beta}{\nu}
\sum_{n~even}{\frac{1}{x^{n+1}}
4\sum_{j=0}^{\infty}{
{\cal C}(n,j)j~d^+_{n+2j}    }}
\nonumber\\
& &
\hspace*{-1cm}
+ \varepsilon_{\mu\alpha\nu\beta}
\frac{q^\alpha P^\beta(S\cdot q)-q^\alpha S^\beta(P\cdot q)}{\nu^2}
\nonumber\\ & & \hspace{2cm} \times
\sum_{n~even}{\frac{1}{x^{n+1}}
(n+1)\sum_{j=0}^{\infty}{
{\cal C}_1(n-1,j)(n+2j)~d^+_{n+2j}  } \bigg\}}
\nonumber\\
& &
\hspace*{-1cm}
- (g_{V_1}g_{A_2}+g_{V_2}g_{A_1})
 \bigg\{(-g_{\mu\nu}+\frac{q_\mu q_\nu}{q^2})\frac{S\cdot q}{\nu} 
\sum_{n~odd}\frac{1}{x^{n+1}}
4\sum_{j=0}^{\infty}{
{\cal C}(n,j)j~\frac{d^+_{n+2j}}{(n+2j)} } 
\nonumber\\
& &
\hspace*{-1cm}
- \biggl(\frac{\hat S^\mu\hat P^\nu+\hat P^\mu\hat S^\nu}{2}-
\frac{\hat P^\mu\hat P^\nu}{\nu}(S\cdot q)\biggr)
\frac{2}{\nu}\sum_{n~odd}{\frac{1}{x^n}
\sum_{j=0}^{\infty}{
{\cal C}_1(n-1,j)d^+_{n+2j}} }\nonumber\\
&&\hspace{5.1cm} \times \left[(n-1)(n+1)+4j(n+2j+1)\right]\nonumber\\
& &
\hspace*{-1cm}
- \frac{\hat P^\mu\hat P^\nu}{\nu}(S\cdot q)
\frac{8}{\nu}\sum_{n~odd}\frac{1}{x^n}
\sum_{j=0}^{\infty}{
{\cal C}_1(n-1,j)(n+j+1)j~d^+_{n+2j}         } \bigg\}~,
\label{TCompNCt3}
\end{eqnarray}
and
\begin{eqnarray}
\langle PS|\widehat T_{\mu\nu, spin}^{-,~\tau=3}|PS\rangle 
&=& -i(g_{V_1}g_{V_2}+g_{A_1}g_{A_2})
\bigg\{
\varepsilon_{\mu\alpha\nu\beta}\frac{q^\alpha S^\beta}{\nu}
\sum_{n~odd}{\frac{1}{x^{n+1}}
4\sum_{j=0}^{\infty}{
{\cal C}(n,j)j~d^+_{n+2j}    }}\nonumber\\
&+&\varepsilon_{\mu\alpha\nu\beta}
\frac{q^\alpha P^\beta(S\cdot q)-q^\alpha S^\beta(P\cdot q)}{\nu^2}
\nonumber\\ & & \hspace{2cm}
\sum_{n~odd}{\frac{1}{x^{n+1}}
(n+1)\sum_{j=0}^{\infty}{
{\cal C}_1(n-1,j)(n+2j)~d^+_{n+2j}  }}
\bigg\}
\nonumber\\
& -&(g_{V_1}g_{A_2}+g_{V_2}g_{A_1})
 \bigg\{(-g_{\mu\nu}+\frac{q_\mu q_\nu}{q^2})\frac{S\cdot q}{\nu} 
\sum_{n~even}\frac{1}{x^{n+1}}
4\sum_{j=0}^{\infty}{
{\cal C}(n,j)j~\frac{d^+_{n+2j}}{(n+2j)} } 
\nonumber\\
&-&\biggl(\frac{\hat S^\mu\hat P^\nu+\hat P^\mu\hat S^\nu}{2}-
\frac{\hat P^\mu\hat P^\nu}{\nu}(S\cdot q)\biggr)
\frac{2}{\nu}\sum_{n~even}{\frac{1}{x^n}
\sum_{j=0}^{\infty}{
{\cal C}_1(n-1,j)d^+_{n+2j}} }\nonumber\\
&&\hspace{6.1cm} \times
\left[(n-1)(n+1)+4j(n+2j+1)\right]\nonumber\\
&-&\frac{\hat P^\mu\hat P^\nu}{\nu}(S\cdot q)
\frac{8}{\nu}\sum_{n~even}\frac{1}{x^n}
\sum_{j=0}^{\infty}{
{\cal C}_1(n-1,j)(n+j+1)j~d^+_{n+2j}         } \bigg\}~,
\label{TComp-t3}
\end{eqnarray}
where the combinatoric factors ${\cal C}(n,j)$ and ${\cal C}_1(n,j)$
are given by
\begin{eqnarray}
{\cal C}(n,j) & = & \left(\frac{M^2}{Q^2}\right)^j\frac{(n+j+1)!}
{j!n!(n+2j+1)^2}~,\\
{\cal C}_1(n,j) & = & \left(\frac{M^2}{Q^2}\right)^j\frac{(n+j+1)!}
{j!n!(n+2j+1)(n+2j+2)^2}~.
\end{eqnarray}
In the presence of target mass effects all polarized structure functions
receive twist--3 contributions. Implicitly this effect was contained
in Ref.~\cite{MU} already in the case of the structure function $g_1$.
\section{Expressions for the Moments of Structure Functions}

\vspace{2mm}
\noindent
To obtain the moments of the twist--2 and the twist--3 parts of the
individual polarized structure functions we first consider the target
mass corrections to the single amplitudes $A_{i, \tau = 2,3}^{NC,+,-}$.
The twist--2 terms are~:
\begin{eqnarray}
A_{1~\tau=2}^{NC,+}(q^2,\nu) &=& \frac{M^2}{\nu}
\sum_{n~even}{\frac{((g_V^q)^2+(g_A^q)^2) }{x^{n+1}}
(n+1)\sum_{j=0}^{\infty}{
{\cal C}(n,j) a^{+q}_{n+2j}}}~,\nonumber\\
A_{2~\tau=2}^{NC,+}(q^2,\nu) &=& -\frac{M^2}{\nu}
\sum_{n~even}{\frac{((g_V^q)^2+(g_A^q)^2)}{x^{n+1}}
n\sum_{j=0}^{\infty}{
{\cal C}(n,j) a^{+q}_{n+2j}}}~,\nonumber\\
A_{3~\tau=2}^{NC,+}(q^2,\nu) &=& \frac{M^2}{\nu}  
\sum_{n~odd}{\frac{8 g_V^q g_A^q}{x^n}
n\sum_{j=0}^{\infty}{
{\cal C}(n,j) \frac{a^{+q}_{n+2j}}{n+2j}   }}~,\nonumber\\
A_{4~\tau=2}^{NC,+}(q^2,\nu) &=& \frac{M^4}{\nu^2}  
\sum_{n~odd}{\frac{4 g_V^q g_A^q}{x^n}
n(n+1)\sum_{j=0}^{\infty}{
{\cal C}(n,j) \frac{a^{+q}_{n+2j}}{n+2j}   }}~,\nonumber\\
A_{5~\tau=2}^{NC,+}(q^2,\nu) &=& \frac{M^4}{\nu^2}   
\sum_{n~odd}{\frac{2 g_V^q g_A^q}{x^{n+1}}
\sum_{j=0}^{\infty}{
{\cal C}(n,j) \frac{a^{+q}_{n+2j}}{n+2j}
[(n+2j+1)(n+2j)+2j]}}~,
\end{eqnarray}
and
\begin{eqnarray}
A_{1~\tau=2}^{-}(q^2,\nu) &=& \frac{M^2}{\nu}
\sum_{n~odd}{\frac{((g_V^q)^2+(g_A^q)^2) }{x^{n+1}}
(n+1)\sum_{j=0}^{\infty}{
{\cal C}(n,j) a^{-q}_{n+2j}}}~,\nonumber\\
A_{2~\tau=2}^{-}(q^2,\nu) &=& -\frac{M^2}{\nu}
\sum_{n~odd}{\frac{((g_V^q)^2+(g_A^q)^2)}{x^{n+1}}
n\sum_{j=0}^{\infty}{
{\cal C}(n,j) a^{-q}_{n+2j}}}~,\nonumber\\
A_{3~\tau=2}^{-}(q^2,\nu) &=& \frac{M^2}{\nu}  
\sum_{n~even}{\frac{8 g_V^q g_A^q}{x^n}
n\sum_{j=0}^{\infty}{
{\cal C}(n,j) \frac{a^{-q}_{n+2j}}{n+2j}   }}~,\nonumber\\
A_{4~\tau=2}^{-}(q^2,\nu) &=& \frac{M^4}{\nu^2}  
\sum_{n~even}{\frac{4 g_V^q g_A^q}{x^n}
n(n+1)\sum_{j=0}^{\infty}{
{\cal C}(n,j) \frac{a^{-q}_{n+2j}}{n+2j}   }}~,\nonumber\\
A_{5~\tau=2}^{-}(q^2,\nu) &=& \frac{M^4}{\nu^2}   
\sum_{n~even}{\frac{2 g_V^q g_A^q}{x^{n+1}}
\sum_{j=0}^{\infty}{
{\cal C}(n,j) \frac{a^{-q}_{n+2j}}{n+2j}
[(n+2j+1)(n+2j)+2j]}}~.
\end{eqnarray}
The twist--3 terms read~:
\begin{eqnarray}
A_{1~\tau=3}^{NC,+}(q^2,\nu) &=& \frac{M^2}{\nu}
\sum_{n~even}{\frac{((g_V^q)^2+(g_A^q)^2) }{x^{n+1}}
4\sum_{j=0}^{\infty}{
{\cal C}(n,j)j d^{+q}_{n+2j}}}~,\nonumber\\
A_{2~\tau=3}^{NC,+}(q^2,\nu) &=& \frac{M^2}{\nu}
\sum_{n~even}{\frac{((g_V^q)^2+(g_A^q)^2)}{x^{n+1}}
\sum_{j=0}^{\infty}{
(n+1){\cal C}_1(n-1,j)(n+2j) d^{+q}_{n+2j}}}~,\nonumber\\
A_{3~\tau=3}^{NC,+}(q^2,\nu) &=& \frac{M^2}{\nu}  
\sum_{n~odd}{\frac{4 g_V^q g_A^q}{x^n}
\sum_{j=0}^{\infty}{
{\cal C}_1(n-1,j)\left[(n-1)(n+1)+4j(n+2j+1)\right]d^{+q}_{n+2j}  }}~,
\nonumber\\
A_{4~\tau=3}^{NC,+}(q^2,\nu) &=& \frac{M^4}{\nu^2}  
\sum_{n~odd}{\frac{16 g_V^q g_A^q}{x^n}
n(n+1)\sum_{j=0}^{\infty}{
{\cal C}_1(n-1,j)(n+j+1)j d^{+q}_{n+2j}   }}~,\nonumber\\
A_{5~\tau=3}^{NC,+}(q^2,\nu) &=& -\frac{M^4}{\nu^2}   
\sum_{n~odd}{\frac{8 g_V^q g_A^q}{x^{n+1}}
\sum_{j=0}^{\infty}{
{\cal C}(n,j) \frac{d^{+q}_{n+2j}}{n+2j}      }}~,
\end{eqnarray}
and
\begin{eqnarray}
A_{1~\tau=3}^{-}(q^2,\nu) &=& \frac{M^2}{\nu}
\sum_{n~odd}{\frac{((g_V^q)^2+(g_A^q)^2) }{x^{n+1}}
4\sum_{j=0}^{\infty}{
{\cal C}(n,j)j d^{+q}_{n+2j}}}~,\nonumber\\
A_{2~\tau=3}^{-}(q^2,\nu) &=& \frac{M^2}{\nu}
\sum_{n~odd}{\frac{((g_V^q)^2+(g_A^q)^2)}{x^{n+1}}
\sum_{j=0}^{\infty}{
(n+1){\cal C}_1(n-1,j)(n+2j) d^{+q}_{n+2j}}}~,\nonumber\\
A_{3~\tau=3}^{-}(q^2,\nu) &=& \frac{M^2}{\nu}  
\sum_{n~even}{\frac{4 g_V^q g_A^q}{x^n}
\sum_{j=0}^{\infty}{
{\cal C}_1(n-1,j)\left[(n-1)(n+1)+4j(n+2j+1)\right]d^{+q}_{n+2j}  }}~,
\nonumber\\
A_{4~\tau=3}^{-}(q^2,\nu) &=& \frac{M^4}{\nu^2}  
\sum_{n~even}{\frac{16 g_V^q g_A^q}{x^n}
n(n+1)\sum_{j=0}^{\infty}{
{\cal C}_1(n-1,j)(n+j+1)j d^{+q}_{n+2j}   }}~,\nonumber\\
A_{5~\tau=3}^{-}(q^2,\nu) &=& -\frac{M^4}{\nu^2}   
\sum_{n~even}{\frac{8 g_V^q g_A^q}{x^{n+1}}
\sum_{j=0}^{\infty}{
{\cal C}(n,j) \frac{d^{+q}_{n+2j}}{n+2j}         }}~.
\end{eqnarray}

From the representations Eqs.~(\ref{tayNC},\ref{tay-}) the following 
relations between the moments of the twist--2 and twist--3 parts of the
structure functions $g_i^{{\rm NC},\pm}(x,Q^2)$ and the operator matrix
elements $a_n^{\pm,q}$ and $d_n^{\pm,q}$ are obtained. The twist--2 
contributions are
\ba
&&\int_0^1dx~x^ng_{1~\tau=2}^{NC,+}(x,Q^2)
=\sum_q\frac{(g_V^q)^2+(g_A^q)^2}{4}
(n+1)\sum_{j=0}^{\infty}{
{\cal C}(n,j)
a^{+q}_{n+2j}}~,~~~n=0,2~...\label{g1NC}\\
&&\int_0^1dx~x^ng_{2~\tau=2}^{NC,+}(x,Q^2)
=-\sum_q\frac{(g_V^q)^2+(g_A^q)^2}{4}
~n~\sum_{j=0}^{\infty}{
{\cal C}(n,j)
a^{+q}_{n+2j}}~,~~~~~n=2,4~...\label{g2NC}\\
&&\int_0^1dx~x^ng_{3~\tau=2}^{NC,+}(x,Q^2)= 
\sum_q{2 g_V^qg_A^q}
(n+1)\sum_{j=0}^{\infty}{
{\cal C}(n+1,j) 
\frac{a^{+q}_{n+2j+1}}{(n+2j+1)}}~,~~~n=0,2~...\label{g3NC}\\
&&\int_0^1dx~x^ng_{4~\tau=2}^{NC,+}(x,Q^2)=
\sum_q {~g_V^qg_A^q} 
(n+1)(n+2)\sum_{j=0}^{\infty}{
{\cal C}(n+1,j) 
\frac{a^{+q}_{n+2j+1}}{(n+2j+1)}}~, \nonumber\\
&&\hspace{12cm} ~~~n=2,4~...\label{g4NC}\\
&&\int_0^1dx~x^ng_{5 \tau=2}^{NC,+}(x,Q^2)=
\sum_q{\frac{1}{2}g_V^qg_A^q}
\sum_{j=0}^{\infty}{
{\cal C}(n,j)
\frac{a^{+q}_{n+2j}}{(n+2j)}}\nonumber\\
&&\hspace{6.5cm} \times [(n+2j+1)(n+2j)+2j] ,~~~n=1,3~...,\label{g5NC}
\ea
and 
\ba
&&\int_0^1dx~x^ng_{1 \tau=2}^{-}(x,Q^2)=\sum_q\frac{(g_V^q)^2
+(g_A^q)^2}{4}
(n+1)\sum_{j=0}^{\infty}{
{\cal C}(n,j)
a^{-q}_{n+2j}}~,~~~n=1,3~...\label{g1-}\\
&&\int_0^1dx~x^ng_{2 \tau=2}^{-}(x,Q^2)=-\sum_q\frac{(g_V^q)^2
+(g_A^q)^2}{4}
~n~\sum_{j=0}^{\infty}{
{\cal C}(n,j)
a^{-q}_{n+2j}}~,~~~~~~n=1,3~...\label{g2-}\\
&&\int_0^1dx~x^ng_{3 \tau=2}^{-}(x,Q^2)= 
\sum_q{2 g_V^qg_A^q}
(n+1)\sum_{j=0}^{\infty}{
{\cal C}(n+1,j) 
\frac{a^{-q}_{n+2j+1}}{(n+2j+1)}}~,~~~n=1,3~...\label{g3-}\\
&&\int_0^1dx~x^ng_{4 \tau=2}^{-}(x,Q^2)=
\sum_q {~g_V^qg_A^q} 
(n+1)(n+2)\sum_{j=0}^{\infty}{
{\cal C}(n+1,j) 
\frac{a^{-q}_{n+2j+1}}{(n+2j+1)}}~,\nonumber\\
&&\hspace{12cm}~~~n=1,3~...\label{g4-}\\
&&\int_0^1dx~x^ng_{5 \tau=2}^{-}(x,Q^2)=
\sum_q{\frac{1}{2}g_V^qg_A^q}
\sum_{j=0}^{\infty}{
{\cal C}(n,j)
\frac{a^{-q}_{n+2j}}{(n+2j)}}\nonumber\\
&&\hspace{6.5cm} \times
[(n+2j+1)(n+2j)+2j] ,~~~n=0,2~...\label{g5-}
\ea
For the twist--3 contributions to the structure functions the moments 
are given by
\ba
&&\int_0^1dx~x^n g_{1~\tau=3} ^{NC,+}(x,Q^2)
=\sum_q\frac{(g_V^q)^2+(g_A^q)^2}{4}
4\sum_{j=0}^{\infty}{
{\cal C}(n,j)j
d^{+q}_{n+2j}}~,~~~~~~~~~~~n=0,2~...\label{g1t3NC}\\
&&\int_0^1dx~x^n g_{2~\tau=3}^{NC,+}(x,Q^2)
=\sum_q\frac{(g_V^q)^2+(g_A^q)^2}{4}
\sum_{j=0}^{\infty}{
{\cal C}_1(n-1,j)(n+2j)(n+1) d^{+q}_{n+2j}}~,~~~~~~~~~~\nonumber\\
&&\hspace{12.2cm}n=2,4~...\label{g2t3NC}\\
&&\int_0^1dx~x^n g_{3~\tau=3}^{NC,+}(x,Q^2) =
\sum_q {g_V^q g_A^q
\sum_{j=0}^{\infty}{{\cal C}_1(n,j) d_{n+2j+1}[(n+2j)(n+2j+2)+4j]}}~,
\nonumber\\
&&\hspace{12.2cm}n=0,2~...\label{g3t3NC}\\
&&\int_0^1dx~x^n g_{4 \tau=3}^{NC,+}(x,Q^2) =
4\sum_q {g_V^q g_A^q
\sum_{j=0}^{\infty}{{\cal C}_1(n,j)(n+j+2)j d_{n+2j+1}}}~,~~~~n=2,4~...
\label{g4t3NC}\\
&&\int_0^1dx~x^n g_{5 \tau=3}^{NC,+}(x,Q^2) =
-2\sum_q {g_V^q g_A^q
\sum_{j=0}^{\infty}{{\cal C}(n,j)j \frac{d_{n+2j}}{n+2j}}}~,
~~~~~~~~~~~~~~~~~~n=1,3~...
\label{g5t3NC}
\end{eqnarray}
and 
\ba
&&\int_0^1dx~x^n g_{1~\tau=3}^{-}(x,Q^2)
=\sum_q\frac{(g_V^q)^2+(g_A^q)^2}{4}
4\sum_{j=0}^{\infty}{
{\cal C}(n,j)j
d^{-q}_{n+2j}}~,~~~~~~~n=1,3~...\label{g1t3-}\\
&&\int_0^1dx~x^n  g_{2~\tau=3}^{-}(x,Q^2)
=\sum_q\frac{(g_V^q)^2+(g_A^q)^2}{4}
\sum_{j=0}^{\infty}{
{\cal C}_1(n-1,j)(n+2j)(n+1) d^{-q}_{n+2j} }~,\nonumber\\
&&\hspace{12cm}n=1,3~...\label{g2t3-}\\
&&\int_0^1dx~x^n g_{3~\tau=3}^{-}(x,Q^2) =
\sum_q {g_V^q g_A^q
\sum_{j=0}^{\infty}{{\cal C}_1(n,j) d_{n+2j+1}[(n+2j)(n+2j+2)+4j]}}~,
\nonumber \\
&&\hspace{12cm}n=1,3~...~~~~~~~\label{g3t3-}\\
&&\int_0^1dx~x^n g_{4~\tau=3}^{-}(x,Q^2) =
4\sum_q {g_V^q g_A^q
\sum_{j=0}^{\infty}{{\cal C}_1(n,j)(n+j+2)j d_{n+2j+1} }}~,~~~~~~n=1,3
\label{g4t3-}\\
&&\int_0^1dx~x^n g_{5~\tau=3}^{-}(x,Q^2) =
-2\sum_q {g_V^q g_A^q
\sum_{j=0}^{\infty}{{\cal C}(n,j)j \frac{d_{n+2j}}{n+2j} }}~,
~~~~~~~~~n=0,2~...
\label{g5t3-}
\ea
Eqs.~(\ref{g1NC}--\ref{g5t3-}) can be used directly in structure function 
analyses based on integer moments. Unlike the massless case the operator
expectation values $a_n^{\pm}$ and $d_n^{\pm}$ do not decouple for a
given spin $n$ in this representation. As will be shown in the subsequent
section, however, the infinite sums in Eqs.~(\ref{g1NC}--\ref{g5t3-}) 
can be related to integrals over one--dimensional partition functions via
an analytic continuation from the integers $n$ to the complex $n$-plane.
Carlson's theorem~\cite{CARLS} assures the uniqueness of the analytic
continuation and the inverse Mellin transform.
\section{The Inverse Mellin Transform}

\vspace{2mm}
\noindent
In many practical applications, as the analysis of experimental data,
the expressions for the moments of the twist--2 and twist--3 
contributions to the deep inelastic scattering structure functions at
integer values of $n$ are less suited than the corresponding $x$--space 
expressions. For this purpose we perform the inverse Mellin transform
of the results derived in the last section and provide integral
representations which can be directly applied to the measured structure
functions, which are given as distributions in $x$ and $Q^2$, to unfold 
the target mass corrections.

We apply a procedure analogous to that used by Georgi and Politzer in
Ref.~\cite{GP} in the case of unpolarized structure functions. We
summarize the different steps for the case of the structure functions
$g_1^{\pm}(x,Q^2)$. From Eqs.~(\ref{g1NC},\ref{g1-}) one obtains
\begin{eqnarray}
g_{1~\tau=2}^{\pm}(x,Q^2) &=& \sum_q \frac{(g_V^q)^2+(g_A^q)^2}{4}
\nonumber\\ & & \times \frac{1}{2\pi i}\int_{-i\infty}^{+i\infty}
dn~x^{-(n+1)}(n+1) \sum_{j=0}^{\infty}{\left(\frac{M^2}{Q^2}\right)^j
\frac{(n+j+1)!}{j!n!} \frac{a_{n+2j}^{\pm,q}}{(n+2j+1)^2}}~.
\label{g1Px}
\end{eqnarray}
The operator expectation values  $a^{\pm}_n$ are the moments of  
distribution functions $F^{\pm q}(x)$, which are related to the
polarized parton densities in the massless limit, $\Delta q(x) \pm
\Delta \overline{q}(x)$, cf. e.g.~\cite{BK},
\begin{eqnarray}
a_n^{\pm,q} = \int_{0}^{1}{dy y^n  F^{\pm q}(y)}~.
\label{an}
\end{eqnarray}
Hence,
\begin{eqnarray}
\frac{a_{n+2j}^\pm}{(n+2j+1)^2} &=&\int_{0}^{1}{dy y^{n+2j} G_1^\pm(y)}~,
\end{eqnarray}
where
\begin{eqnarray}
G^\pm_1(y) &=& 
\sum_q\frac{(g_V^q)^2+(g_A^q)^2}{4}
\int_{y}^{1}{\frac{dy_1}{y_1}\int_{y_1}^{1}{\frac{dy_2}{y_2}F^{\pm q}(y_2)}}~.
\end{eqnarray}
Inserting these expressions into Eq.~(\ref{g1Px}) and interchanging the 
integrations and the summation, the following expression is obtained
\begin{eqnarray}
g_{1~\tau=2}^+(x)& = &\int_{0}^{1}{dy G^{\pm}_1(y)
 \frac{1}{2\pi i}\int_{-i\infty}^{+i\infty} 
dn~ y^n x^{-(n+1)}(n+1)^2 
\sum_{j=0}^{\infty}{\left(\frac{M^2 y^2}{Q^2}\right)^j
\frac{(n+j+1)!}{j!(n+1)!}}}~.
\end{eqnarray}
The infinite sum over $j$ in this equation is performed by observing that
\renewcommand{\arraystretch}{1}
\begin{eqnarray}
\frac{1}{(1-r)^{n+1}}& =& \sum_{j=0}^{\infty}{\frac{(n+j)!}{j!n!}r^j}
= \sum_{l,k \geq 0} 
\left[\begin{array}{c}k\\l\end{array}\right] (n+1)^l
\frac{r^k}{k!}~.
\end{eqnarray}
This relation can be derived by differentiating the generating
functional of the geometric series, $(1-r)^{-1}$. Here,
$\left[\begin{array}{c}k\\l\end{array}\right]$ denotes the Stirling
numbers of the first kind, cf.~\cite{GKP}.

The integration variable $n$ is brought into exponential from by
introducing differential operators  w.r.t. $x$
\begin{eqnarray}
g_{1~\tau=2}^{\pm}(x)& = &
x\frac{d}{ dx} x\frac{ d}{d x}
\int_{0}^{1}{dy \frac{G^{\pm}_1(y)}
{x\left(1-M^2 y^2/Q^2\right)}
 \frac{1}{2\pi i}\int_{-i\infty}^{+i\infty} 
dn y^n x^{-n}
\left(1-\frac{M^2 y^2}{Q^2}\right)^{-n}}.
\end{eqnarray}
The $n$--integration leads to a $\delta$-function
\begin{eqnarray}
g_{1~\tau=2}^{\pm}(x)& = &
x\frac{d}{ dx} x\frac{ d}{d x}
 \int_{0}^{1}{dy {\ds \frac{G^{\pm}_1(y)}{x 
 \left(1-\ds
 \frac{
 M^2 y^2}{Q^2}\right)}}
\delta\left[\ln(y)-\ln(x)-\ln\left(1-\frac{M^2 y^2}{Q^2}\right)
\right]}~.
\end{eqnarray}
Finally one obtains
\begin{eqnarray}
g_{1~\tau=2}^{\pm}(x)& = &
x\frac{d}{ dx} x\frac{ d}{d x}
\left[\frac{x}{(1+4M^2 x^2/Q^2)^{1/2}} 
\frac{G^{\pm}_1(\xi)}{\xi}\right]~.
\label{g1G}
\end{eqnarray}
Here, $\xi$ denotes the Nachtmann--variable~\cite{Nachtmann},
\begin{eqnarray}
\xi &=& \frac{2x}{1+(1+4M^2 x^2/Q^2)^{1/2}}~.
\end{eqnarray}
Similar expressions are obtained for the other structure functions
\begin{eqnarray}
g_{2~\tau=2}^{\pm}(x)& = &-
x\frac{d^2}{ dx^2}x
\left[\frac{x}{(1+4M^2 x^2/Q^2)^{1/2}} \frac{G^{\pm}_1(\xi)}{\xi}
\right]~,
\label{g2G}\\
g_{3~\tau=2}^{\pm}(x) & = & 
2 x^2\frac{d^2}{ dx^2}
\left[\frac{x^2}{(1+4M^2 x^2/Q^2)^{1/2}} \frac{G^{\pm}_2(\xi)}{\xi^2}
\right]~,
\label{g3G}\\
g_{4~\tau=2}^{\pm}(x) & = & -
x^2\frac{d}{ dx}x\frac{d^2}{dx^2}
\left[\frac{x^2}{(1+4M^2 x^2/Q^2)^{1/2}} \frac{G^{\pm}_2(\xi)}{\xi^2}
\right]~,
\label{g4G}\\
g_{5~\tau=2}^{\pm}(x) & = & -
x\frac{d}{ dx}
\left[\frac{x}{(1+4M^2 x^2/Q^2)^{1/2}} \frac{G^{\pm}_3(\xi)}{\xi}\right]
+\frac{M^2}{Q^2}x^2\frac{d^2}{ dx^2}
\left[\frac{x^2}{(1+4M^2 x^2/Q^2)^{1/2}} \frac{G^{\pm}_2(\xi)}{\xi}
\right]~. \nonumber\\
\label{g5G}
\end{eqnarray}
The functions $G^{\pm}_2(y)$ and $G^{\pm}_3(y)$ are related to the
distribution function $F^{\pm q}(y)$ by
\begin{eqnarray}
G^{\pm}_2(y) & = &\sum_q {~g_V^qg_A^q 
\int_{y}^{1}{dy_1\int_{y_1}^{1}{\frac{dy_2}{y_2}{\int_{y_2}^{1}
{\frac{dy_3}{y_3}F^{\pm q}(y_3)}}}}}~,\\
G^{\pm}_3(y) & = & \sum_q {\frac{1}{2}~g_V^qg_A^q
\int_{y}^{1}{\frac{dy_1}{y_1}F^{\pm q}(y_1)}}~.
\end{eqnarray}
Performing the derivatives in Eqs.~(\ref{g1G}, \ref{g2G}--\ref{g5G}), the
twist--2 contributions to the structure functions can be expressed in 
terms of the distribution functions $F^{\pm q}(\xi)$ as follows~:
\begin{eqnarray}
g_{1~\tau=2}^{\pm}(x) 
&=& 
\sum_q \frac{(g_V^q)^2+(g_A^q)^2}{4} \Biggl\{
\frac{x}{\xi}
\frac{F^{\pm q}(\xi)}{(1+4M^2 x^2/Q^2)^{3/2}}
\nonumber\\
&+& \frac{4 M^2 x^2}{Q^2}
\frac{(x+\xi)}{\xi(1+4M^2 x^2/Q^2)^2}
\int_{\xi}^{1}\frac{d\xi_1}{\xi_1} F^{\pm q}(\xi_1)
\nonumber \\
&-&\frac{4 M^2 x^2}{Q^2} 
\frac{(2-4 M^2 x^2/Q^2)}{2(1+4 M^2 x^2/Q^2)^{5/2}}
\int_{\xi}^{1}{\frac{d\xi_1}{\xi_1}\int_{\xi_1}^{1}{
\frac{d\xi_2}{\xi_2}F^{\pm q}(\xi_2)}}
\Biggr\}~,
\label{g1x}\\
g_{2~\tau=2}^{\pm}(x)
&=&
\sum_q \frac{(g_V^q)^2+(g_A^q)^2}{4}\Biggl\{
 -\frac{x}{\xi}\frac{F^{\pm q}(\xi)}{(1+4M^2 x^2/Q^2)^{3/2}}
 \nonumber\\
&+&  
\frac{x(1-4 M^2 x\xi/Q^2)}{\xi(1+4M^2 x^2/Q^2)^2}
\int_{\xi}^{1}\frac{d\xi_1}{\xi_1} F^{\pm q}(\xi_1) 
\nonumber \\
&+&\frac{3}{2}
\frac{4 M^2 x^2/Q^2}{(1+4 M^2 x^2/Q^2)^{5/2}}
\int_{\xi}^{1}{\frac{d\xi_1}{\xi_1}\int_{\xi_1}^{1}{
\frac{d\xi_2}{\xi_2}F^{\pm q}(\xi_2)}}\Biggr\}~,
\label{g2x}\\
g_{3~\tau=2}^{\pm}(x) &=&
\sum_q {~g_V^qg_A^q 
\Biggl\{
\frac{2x^2}{\xi(1+4 M^2 x^2/Q^2)^{3/2}}
\int_{\xi}^{1}{\frac{d\xi_1}{\xi_1}F^{\pm q}(\xi_1)}}\nonumber\\  
&+&\frac{12 M^2 x^3/Q^2}{(1+4 M^2 x^2/Q^2)^2}
\int_{\xi}^{1}{\frac{d\xi_1}{\xi_1}\int_{\xi_1}^{1}{
\frac{d\xi_2}{\xi_2}F^{\pm q}(\xi_2)}}\nonumber \\
&+&\frac{3}{2}\frac{(4 M^2 x^2/Q^2)^2}{(1+ 4 M^2 x^2/Q^2)^{5/2}}
\int_{\xi}^{1}{d\xi_1\int_{\xi_1}^{1}{
\frac{d\xi_2}{\xi_2}\int_{\xi_2}^{1}{\frac{d\xi_3}{\xi_3}F^{\pm q}
(\xi_3)}}}
\Biggr\}~,
\label{g3x}\\
g_{4~\tau=2}^{\pm}(x) &=&
\sum_q {~g_V^qg_A^q
\Biggl\{
\frac{x^2}{\xi}\frac{F^{\pm q}(\xi)}{(1+4 M^2 x^2/Q^2)^2}
+\frac{8 M^2 x^2}{Q^2}
\frac{x (x+\xi)}{\xi(1+4 M^2 x^2/Q^2)^3}
\int_{\xi}^{1}{\frac{d\xi_1}{\xi_1}F^{\pm q}(\xi_1)}} \nonumber\\
&-& 3 \frac{4 M^2 x^3}{Q^2}
\frac{(1- M^2 x(4x+\xi)/Q^2)}{(1+4 M^2 x^2/Q^2)^3}
\int_{\xi}^{1}{\frac{d\xi_1}{\xi_1}\int_{\xi_1}^{1}{
\frac{d\xi_2}{\xi_2}F^{\pm q}(\xi_2)}}\nonumber \\
&-&\frac{3}{4}\frac{(4 M^2 x^2/Q^2)^2(3-8M^2x^2/Q^2)}
{(1+ 4 M^2 x^2/Q^2)^{7/2}}
\int_{\xi}^{1}{d\xi_1\int_{\xi_1}^{1}{
\frac{d\xi_2}{\xi_2}\int_{\xi_2}^{1}{\frac{d\xi_3}{\xi_3}F^{\pm q}
(\xi_3)}}}
\Biggr\}~,
\label{g4x}\\
g_{5~\tau=2}^{\pm}(x)  &=&
\sum_q{\frac{1}{2}~g_V^qg_A^q  \Biggl\{
\frac{x}{\xi(1+4 M^2 x^2/Q^2)}F^{\pm q}(\xi)} \nonumber\\
&+& {
 2\frac{M^2x^2}{Q^2}\Biggl( 
\frac{(1+\xi)}{\xi(1+4 M^2 x^2/Q^2)^{3/2}}
\int_{\xi}^{1}{\frac{d\xi_1}{\xi_1}F^{\pm q}(\xi_1)}}\nonumber\\
&+&\frac{2(x-2\xi)}{\xi(1+4 M^2 x^2/Q^2)^2}
\int_{\xi}^{1}{\frac{d\xi_1}{\xi_1}\int_{\xi_1}^{1}{
\frac{d\xi_2}{\xi_2}F^{\pm q}(\xi_2)}}\nonumber\\
&-&\frac{6M^2x/Q^2}{(1+ 4 M^2 x^2/Q^2)^{5/2}}
\int_{\xi}^{1}{d\xi_1\int_{\xi_1}^{1}{
\frac{d\xi_2}{\xi_2}\int_{\xi_2}^{1}{\frac{d\xi_3}{\xi_3}F^{\pm q}
(\xi_3)}}}
\Biggr)\Biggr\}~.
\label{g5x}
\end{eqnarray} 

In the same way the moments of the corresponding twist--3 contributions,
Eqs.~(\ref{g1t3NC}--\ref{g5t3-}), are inverted. One obtains the
following $x$-space expressions~:
\begin{eqnarray}
 g^{\pm}_{1~\tau=3}(x,Q^2) &=& \frac{4 M^2}{Q^2} x^2\frac{d^2}{dx^2}
\Biggl[\frac{x^2}{(1+4M^2x^2/Q^2)^{1/2}} H^{\pm}_1(\xi)\Biggr]~,
\label{g1t3dif}\\
 g^{\pm}_{2~\tau=3}(x,Q^2) &=&  x\frac{d^2}{dx^2}
\Biggl[\frac{x}{(1+4M^2x^2/Q^2)^{1/2}} H^{\pm}_1(\xi)\Biggr]~,
\label{g2t3dif}\\
g^{\pm}_{3~\tau=3}(x,Q^2) &=& 
-\left(x^2\frac{d^3}{dx^3}+4\frac{M^2x^2}{Q^2}x\frac{d^2}{dx^2}x\right)
\left[\frac{x^2}{(1+4 M^2x^2/Q^2)^{1/2}}
\frac{H_2^{\pm}(\xi)}{\xi}\right]~,
\label{g3t3dif}\\
 g^{\pm}_{4~\tau=3}(x,Q^2) &=& -4 \frac{M^2}{Q^2}
x^3\frac{d^3}{dx^3}
\Biggl[\frac{x^3}{(1+4M^2x^2/Q^2)^{1/2}} \frac{H^{\pm}_2(\xi)}{\xi}
\Biggr]~,
\label{g4t3dif}\\
 g^{\pm}_{5~\tau=3}(x,Q^2) &=& -2 \frac{M^2}{Q^2}
x^2\frac{d^2}{dx^2}
\Biggl[\frac{x^2}{(1+4M^2x^2/Q^2)^{1/2}} \frac{H^{\pm}_2(\xi)}{\xi}
\Biggr]~.
\label{g5t3dif}
\end{eqnarray} 
Here we define
\begin{eqnarray}
H^{\pm}_1(y) &=& 
\sum_q\frac{(g_V^q)^2+(g_A^q)^2}{4}
\int_{y}^{1}{\frac{dy_1}{y_1}\int_{y_1}^{1}{\frac{dy_2}{y_2}
D^{\pm q}(y_2)}}~,
\\
H^{\pm}_2(y) &=&
\sum_q{g_V^q g_A^q
\int_{y}^{1} {dy_1{
\int_{y_1}^{1}{\frac{dy_2}{y_2}
\int_{y_2}^{1}{\frac{dy_3}{y_3}D^{\pm q}(y_3)}}}}}~.
\end{eqnarray}
The matrix elements of the twist--3 operators are the moments of the
distribution function $D^{\pm q}(x)$ in the massless limit, 
\begin{eqnarray}
d^{\pm q}_n=\int_{0}^{1}dx x^n D^{\pm q}(x)~,
\end{eqnarray}
which has, however, no partonic interpretation. Performing the derivatives
in Eqs.~(\ref{g1t3dif}--\ref{g5t3dif}) the following representation
for the twist--3 parts of the polarized structure functions are
obtained~:
\begin{eqnarray}
g_{1~\tau=3}^{\pm}(x,Q^2) &=&
\sum_q{\frac{(g_V^q)^2+(g_A^q)^2}{4} \frac{4M^2x^2}{Q^2}\Biggl\{
 \frac{D^{\pm q}(\xi)}{(1+4M^2x^2/Q^2)^{3/2}}}\nonumber\\
&-& 
\frac{3}{(1+4M^2x^2/Q^2)^2}
\int_{\xi}^{1}{\frac{d\xi_1}{\xi_1} D^{q}(\xi_1)}
\nonumber\\
&+&
\frac{(2-4M^2 x^2/Q^2)}{(1+4M^2x^2/Q^2)^{5/2}}
\int_{\xi}^{1}{\frac{d\xi_1}{\xi_1}\int_{\xi_1}^{1}{
\frac{d\xi_2}{\xi_2}D^{\pm q}(\xi_2)}}\Biggr\}~,
\label{g1t3x}\\
g_{2~\tau=3}^{\pm}(x,Q^2) &=&
\sum_q{\frac{(g_V^q)^2+(g_A^q)^2}{4} \Biggl\{
 \frac{D^{\pm q}(\xi)}{(1+4M^2x^2/Q^2)^{3/2}}}\nonumber\\
&-&\frac{1-8M^2x^2/Q^2}{(1+4M^2x^2/Q^2)^2}
\int_{\xi}^{1}{\frac{d\xi_1}{\xi_1} D^{q}(\xi_1)}
\nonumber\\
&-&
\frac{12M^2x^2/Q^2}{(1+4M^2x^2/Q^2)^{5/2}}
\int_{\xi}^{1}{\frac{d\xi_1}{\xi_1}\int_{\xi_1}^{1}{
\frac{d\xi_2}{\xi_2}D^{\pm q}(\xi_2)}}\Biggr\}~,
\label{g2t3x}\\
g_{3~\tau=3}^{\pm}(x,Q^2) &=&
\sum_q{g_V^q g_A^q \Biggl\{
\frac{x D^{\pm q}(\xi)}{(1+4M^2x^2/Q^2)}
-\frac{2x(1-M^2x\xi/Q^2)}{(1+4M^2x^2/Q^2)^{3/2}}
\int_{\xi}^{1}{\frac{d\xi_1}{\xi_1} D^{\pm q}(\xi_1)}}\nonumber\\
&-&6\frac{M^2 x^2}{Q^2}\frac{(2x+\xi)}{(1+4M^2x^2/Q^2)^2}
\int_{\xi}^{1}{\frac{d\xi_1}{\xi_1}\int_{\xi_1}^{1}
{\frac{d\xi_2}{\xi_2}D^{\pm q}(\xi_2)}}\nonumber\\
&+&
6\frac{M^2x^2}{Q^2}\frac{(1-4M^2x^2/Q^2)}{(1+4M^2x^2/Q^2)^{5/2}}
\int_{\xi}^{1}{d\xi_1\int_{\xi_1}^{1}{\frac{d\xi_2}{\xi_2}
\int_{\xi_2}^{1}
{\frac{d\xi_3}{\xi_3}D^{\pm q}(\xi_3)}}}
\Biggr\}~,
\label{g3t3x}\\
g_{4~\tau=3}^{\pm}(x,Q^2) &=&
-\sum_q{g_V^q g_A^q \frac{4 M^2x^2}{Q^2}\Biggl\{
\frac{x D^{\pm q}(\xi)}{(1+4M^2x^2/Q^2)^2}}\nonumber \\
&+&\frac{x^2(5-2M^2x\xi/Q^2)}{(1+4M^2x^2/Q^2)^{5/2}}
\int_{\xi}^{1}{\frac{d\xi_1}{\xi_1} D^{\pm q}(\xi_1)}
\nonumber\\
&+&
\frac{6 x^2(2 x^2-3\xi^3)}{\xi^2(1+4M^2x^2/Q^2)^3}
\int_{\xi}^{1}{\frac{d\xi_1}{\xi_1}\int_{\xi_1}^{1}
{\frac{d\xi_2}{\xi_2}D^{\pm q}(\xi_2)}}\nonumber\\
&-&
24\frac{M^2x^2}{Q^2}\frac{(1- M^2 x^2/Q^2)}{(1+4M^2x^2/Q^2)^{7/2}}
\int_{\xi}^{1}{d\xi_1\int_{\xi_1}^{1}{\frac{d\xi_2}{\xi_2}
\int_{\xi_2}^{1}
{\frac{d\xi_3}{\xi_3}D^{\pm q}(\xi_3)}}} 
\Biggr\}~,
\label{g4t3x}
\\
g_{5~\tau=3}^{\pm}(x,Q^2) &=&
-2\sum_q{g_V^q g_A^q \frac{ M^2x^2}{Q^2}\Biggl\{
\frac{1}{(1+4M^2x^2/Q^2)^{3/2}}
\int_{\xi}^{1}{\frac{d\xi_1}{\xi_1} D^{\pm q}(\xi_1)} }\nonumber\\
&-&
\frac{2 (1-M^2 x\xi/Q^2)}{(1+4M^2x^2/Q^2)^2}
\int_{\xi}^{1}{\frac{d\xi_1}{\xi_1}\int_{\xi_1}^{1}
{\frac{d\xi_2}{\xi_2}D^{\pm q}(\xi_2)}}\nonumber\\
&-&
\frac{6 M^2x/Q^2}{(1+4M^2x^2/Q^2)^{5/2}}
\int_{\xi}^{1}{d\xi_1\int_{\xi_1}^{1}{\frac{d\xi_2}{\xi_2}
\int_{\xi_2}^{1}
{\frac{d\xi_3}{\xi_3}D^{\pm q}(\xi_3)}}} 
\Biggr\}~.
\label{g5t3x}
\end{eqnarray}
The multiple integrals in the above expressions may be simplified
further to single integrals by using the idendities
\begin{eqnarray}
\int_{\xi}^1 \frac{d\xi_1}{\xi_1} \int_{\xi_1}^1
\frac{d\xi_2}{\xi_2} \phi(\xi_2) &=& \int_{\xi}^1 \frac{d\xi_1}{\xi_1}
\log\left(\frac{\xi_1}{\xi} \right)
\phi(\xi_1)~,\\
\int_{\xi}^1 d\xi_1 \int_{\xi_1}^1
\frac{d\xi_2}{\xi_2} \phi(\xi_2) \int_{\xi_2}^1
\frac{d\xi_3}{\xi_3} \phi(\xi_3) &=& \int_{\xi}^1 \frac{d\xi_1}{\xi_1}
\left[ \xi_1 - \xi - \xi \log \left( \frac{\xi_1}{\xi} \right) \right]
\phi(\xi_1)~.
\end{eqnarray}

We would like to comment on the behavior of 
Eqs.~(\ref{g1x}--\ref{g5x},\ref{g1t3x}--\ref{g5t3x}) 
in the range of
large values of $x$. At first sight these relations for the structure 
functions are not correct. The structure functions should vanish at 
$x=1$, which is not the case for the r.h.s. of
Eqs.~(\ref{g1x}--\ref{g5x},\ref{g1t3x}--\ref{g5t3x}). The inconsistency 
of these equations can be shown also by simply resumming the 
$(M^2/Q^2)$--terms in Eqs. (\ref{g1NC}--\ref{g5t3-}), 
see also \cite{MSG}. As an example, for the moments of $g^+_1(x,Q^2)$ 
one obtains
\begin{eqnarray}
\int_{0}^{1}{dx\,x^n g^+_1(x,Q^2)}& = &
\int_{0}^{1/(1-M^2/Q^2)} {dx\,x^{n+1}
\frac{d}{dx}x\frac{d}{dx}
\left[\frac{x}{(1+4M^2 x^2/Q^2)^{1/2}} \frac{G^+_1(\xi)}{\xi}\right]}.
\label{g1ex}
\end{eqnarray}
The integrand in the r.h.s. of Eq. (\ref{g1ex}) is just the expression of
$g^+_1(x,Q^2)$ after the inverse Mellin transform,  Eq. (\ref{g1x}).
Therefore, the $x$--space expression for $g^+_1(x,Q^2)$, Eq. (\ref{g1x}),
cannot be completely correct because  of the different ranges of 
integration.

This problem can be resolved if we assume that $F(\xi)$, and consequently
$G_1(\xi)$,  vanishes for $\xi>\xi_{th}$, where $\xi_{th}=\xi(x=1)$. 
This assumption is equivalent to the introduction of a kinematic 
threshold factor~\cite{XI} in the moments of distribution functions, as 
was proposed in Ref.~\cite{BJT}. Moreover, only in this case both the 
methods of Refs.~\cite{GP,Nachtmann} are consistent. The matrix elements 
of operators calculated by Nachtmann moments of the structure functions 
(\ref{g1x}--\ref{g5x}) coincide with the definition of the matrix 
elements through the moments of the function $F(x)$, Eq.~(\ref{an}), as 
was observed for the unpolarized structure functions in  Ref.~\cite{DDOR}.
In the range of large values of $x$ the contributions due to dynamical
higher twist operators are increasingly important. As well--known
the number of higher twist operators grows with the spin index $n$ the
more the larger the twist $\tau$~\footnote{The number of these operators
$N(n,\tau)$ is a complicated function in most field theories, cf. e.g.
\cite{HT1,HT3}.} and contribute significantly because of their number
as $x \rightarrow 1$. Therefore the moments of the distribution function
$F(x)$ cannot be expressed by the matrix elements of the lowest twist 
operators only\footnote{Quite a variety of empirical ans\"atze were tried
in the literature to model higher twist contributions to structure 
functions, see~\cite{EMP1,EMP2} and references therein. They are strongly
correlated to the parameters of the lowest twist 
distributions~\cite{EMP2}. These terms are not derived in QCD. Due to the
structure of higher twist operators and coefficient functions, 
cf.~Refs.~[5--7], it is expected that the structure of the higher twist 
contributions is even more complicated than assumed in the ans\"atze 
quoted above.}.
Conversely, the matrix element of the operators in the 
expressions of the moments, Eqs.~(\ref{g1NC}--\ref{g5-}), cannot be
approximated by Eq.~(\ref{an}) alone. As a consequence it turns out, 
that in the approximation considered the results for the structure 
functions in $x$-space are reliable only at $\xi \ll \xi_{th}$.

After having performed the resummation of the $M^2/Q^2$ effects in
Eqs.~(\ref{g1x}--\ref{g5x},\ref{g1t3x}--\ref{g5t3x}) one may try a 
perturbative treatment if $M^2/Q^2 \ll 1$ by expanding these expressions 
into Taylor series. Let us consider the principle relations for the 
structure function $g_1(x,Q^2)$ as an example. Its moments are given by
\begin{eqnarray}
{\cal M}\left[g_1\right](N,Q^2) = \int_0^1 dx x^n g_1(x,Q^2)~.
\end{eqnarray}
The first terms of the corresponding Taylor series are
\begin{eqnarray}
g_1^{n} &= & g_{10}^n +\frac{M^2}{Q^2}
\Bigg\{\frac{(n+1)(n+2)(n+5)}{(n+3)^2}g_{10}^{n+2}
+4\frac{n+1}{n+3}g_{20}^{n+2}\Bigg\}\nonumber\\
&+&\left(\frac{M^2}{Q^2}\right)^2
\Bigg\{\frac{(n+1)(n+2)(n+3)(n+9)}{2(n+5)^2}g_{10}^{n+4}
+4\frac{(n+1)(n+2)(n+3)}{(n+4)(n+5)}g_{20}^{n+4}\Bigg\}\nonumber\\
&+&\left(\frac{M^2}{Q^2}\right)^3
\Biggl\{\frac{(n+1)(n+2)(n+3)(n+4)(n+13)}{6(n+7)^2}g_{10}^{n+6}
\nonumber\\
&&\hspace{4.7cm}+4\frac{(n+1)(n+2)(n+3)(n+4)}{2(n+6)(n+7)}
g_{20}^{n+6}\Bigg\}
+{\cal O}\left(\frac{M^8}{Q^8}\right),
\end{eqnarray}
Here $g^{n}_{10}$ and $g^{n}_{20}$ are the moments of the corresponding
structure functions in the limit $M \rightarrow 0$, with
\begin{eqnarray}
g_{10}^n &=& a_n~,\nonumber\\
g_{20}^n & =& \frac{n}{n+1}(d_n-a_n)~.
\end{eqnarray}
Let us consider furthermore the twist--2 part of $g_1(x)$ 
only,\footnote{The subsequent arguments are similar in the case of the
twist--3 contributions.}
\begin{eqnarray}
g_1^{n} &= & g_{10}^n +\frac{M^2}{Q^2}
\Bigg\{\frac{(n+1)^2(n+2)}{(n+3)^2}g_{10}^{n+2}\Bigg\}\nonumber\\
&+&\left(\frac{M^2}{Q^2}\right)^2
\Bigg\{\frac{(n+1)^2(n+2)(n+3)}{2(n+5)^2}g_{10}^{n+4}\Bigg\}\nonumber\\
&+&\left(\frac{M^2}{Q^2}\right)^3
\Biggl\{\frac{(n+1)^2(n+2)(n+3)(n+4)}{6(n+7)^2}g_{10}^{n+6}\Bigg\}
+{\cal O}\left(\frac{M^8}{Q^8}\right)~.
\label{g1twist2}
\end{eqnarray}
The inverse Mellin transforms of the prefactors of $g_{10}^{n+2}$ and
$g_{10}^{n+4}$ in Eq. (\ref{g1twist2}) are
\begin{eqnarray}
{\cal M}^{-1}\left[\frac{(n+1)^2(n+2)}{(n+3)^2}\right](x)
& = &
-\left(x^3\frac{\rm d}{{\rm d}x}+5x^2\right)\delta(1-x)+4x^2\ln{x}
+8x^2~,\\
{\cal M}^{-1}\left[\frac{(n+1)^2(n+2)(n+3)}{(n+5)^2}\right](x)
& = &
\left(x^5\frac{\rm d^2}{{\rm d}x^2}+12x^5\frac{\rm d}{{\rm d}x}
+64x^4\right)\delta(1-x)
\nonumber \\
&&\hspace{2.8cm} -96x^4\ln{x}-128x^4~,~{\rm etc.}
\end{eqnarray}
One observes that the higher the exponent $k$ of $(M^2/Q^2)^k$ the higher
the order of the  derivatives of the $\delta$--function in the inverse
Mellin transforms given above. Let us assume that the large $x$ behavior
of $g_{10}(x)$ is typically being described by
\begin{eqnarray}
g_{10}(x) \propto (1-x)^a~,
\end{eqnarray}
with $a > 0$. The target mass effects, if represented in terms of a
Taylor expansion, yield
\begin{eqnarray}
g_1(x,Q^2) & \simeq & (1-x)^a 
+ \frac{M^2}{Q^2}\left\{ax(1-x)^{a-1}-5(1-x)^a+I_1(x)\right\}\\
&+&\frac{M^4}{Q^4}\left\{\frac{1}{2}a(a-1)x^2(1-x)^{a-2}
-7ax(1-x)^{a-1}+31(1-x)^a+I_2(x)\right\}+\ldots~, \nonumber
\end{eqnarray}
The integrals
\begin{eqnarray}
I_1(x) &=& 4\int_{x}^{1}{\frac{dx_1}{x_1}\left[2+\ln(x_1)\right]
\left(1-\frac{x}{x_1}\right)^a}~,\nonumber\\
I_2(x) &=& -16\int_{x}^{1}{\frac{dx_1}{x_1}\left[4+3\ln(x_1)\right]
\left(1-\frac{x}{x_1}\right)^a}
\end{eqnarray}
are regular. However, starting at $O(K)$ with
$K > a$ contributions $\propto 1/(1-x)^{|K-a|}$ emerge, which diverge
as $x \rightarrow 1$.  The resummed expressions, on the other hand,
are convergent.
\section{Relations between the Structure Functions~: Twist 2}

\vspace{2mm}
\noindent
In previous analyses three independent
relations between the twist--2 parts of the 
polarized structure functions were derived in the lowest order of the 
coupling constant, the Dicus--relation~\cite{DIC}, the Wandzura--Wilczek
relation~\cite{WW}, and a third relation given in Ref.~\cite{BK}.
We investigate in the following the impact of the target mass corrections
on these relations.

The expressions for the twist--2 contributions to the structure functions
$g_1(x,Q^2)$ and $g_2(x,Q^2)$, Eqs.~(\ref{g1G}, \ref{g2G}), can be 
rewritten in the following form
\begin{eqnarray}
g_1(x,Q^2)&=&x\frac{d}{dx}{\cal F}(x,Q^2)+x^2\frac{d^2}{dx^2}
{\cal F}(x,Q^2)~,
\label{WWg1}\\ 
g_2(x,Q^2)&=&-2x\frac{d}{dx}{\cal F}(x,Q^2)-x^2\frac{d^2}{dx^2}
{\cal F}(x,Q^2)~.
\label{WWg2} 
\end{eqnarray}
Here  ${\cal F}(x,Q^2)$ is given by
\begin{eqnarray}
{\cal F}(x,Q^2) = 
{\ds \frac{x}{\sqrt{1 +  {\ds \frac{4 M^2 x^2}{Q^2}}}}}
\frac{G_1(\xi)}{\xi}~.
\end{eqnarray}
From Eqs.~(\ref{WWg1}, \ref{WWg2}) one obtains
\begin{eqnarray}
g_2(x,Q^2)&=&-g_1(x,Q^2)-x\frac{d}{dx}{\cal F}(x,Q^2)~, 
\label{WWaux1}
\end{eqnarray}
and
\begin{eqnarray}
x\frac{d}{dx}{\cal F}(x,Q^2)&= &-\int_{x}^{1}{\frac{dy}{y}g_1(y,Q^2)}~,
\label{WWaux2}
\end{eqnarray}
from which the Wandzura--Wilczek relation~\cite{WW}
\begin{eqnarray}
g_2(x,Q^2)&= &-g_1(x,Q^2)+\int_{x}^{1}{\frac{dy}{y}g_1(y,Q^2)}
\label{WW}
\end{eqnarray}
results. Therefore the Wandzura--Wilczek relation holds in the
presence of target mass corrections. This result was obtained in 
Ref.~\cite{PR} before. The $Q^2$--behavior in Eq.~(\ref{WW}) refers to the
$(M^2/Q^2)$--corrections in all orders. From 
Eqs.~(\ref{g1NC}, \ref{g2NC}, \ref{g1-}, \ref{g2-}) one obtains
\be
n\int_0^1dx x^ng_1^{i}(x,Q^2) = -(n+1)\int_0^1dx x^n g_2^{i}(x,Q^2),
\label{WW1}
\ee
where $ n=2,4... $ for $i = NC,+$ and $n=1,3...$ for $i = -$. Note that
the moments in Eq.~(\ref{WW1}) are Cornwall--Norton moments, \cite{CN}.
Moreover, it can be shown that the Wandzura--Wilczek relation is also 
valid for the quarkonic operators even in the massive quark case, see 
appendix.  
                                          
The expression for $g_{2, \tau=2}(x,Q^2)$, Eq.~(\ref{WW1}), is formally
consistent
with the Burkhardt--Cottingham sum rule~\cite{BC} in the presence of 
target mass corrections
\begin{eqnarray}
\label{e155}
\int_0^1 dx g^i_{2, \tau = 2}(x,Q^2)=0~,
\end{eqnarray}
although the $0th$ moment of the structure functions $g_2^i(x,Q^2)$ is 
not described by the local operator product expansion.
The same holds for the $0th$ moment of the twist--3 contribution to
$g_2^i(x,Q^2)$, cf.~Eqs.~(\ref{g2t3NC},\ref{g2t3-}).

The relation between the twist--2 contribution to the structure functions
$g_3(x)$ and $g_4(x)$ obtained in Ref.~\cite{BK} in the massless limit
is also valid at any order of $M^2/Q^2$. To show this we integrate
Eq.~(\ref{g4G}) which results into
\be
\label{e300}
g_3^i(x,Q^2)=2x\int_x^1\frac{dy}{y^2}g_4^i(y,Q^2)~.
\ee
This relation can also be  derived directly from the relation between 
the moments of the twist--2 parts of the structure functions $g_3(x)$ and
$g_4(x)$ 
\be
\int_0^1dxx^ng_4^{i}(x,Q^2) = 
\frac{n+2}{2}\int_0^1dxx^ng_3^{i}(x,Q^2)~,
\label{BK1}
\ee
where $ n=2,4... $ for $i = NC, +$ and $n = 1,3...$ for the $i = -$.

In the limit $M \rightarrow 0$  
Eqs.~(\ref{g4NC}, \ref{g5NC}, \ref{g4-}, \ref{g5-}) result into the 
Dicus--relation 
\be
\int_0^1dxx^ng_4^{i}(x,Q^2)=2\int_0^1dxx^{n+1}g_5^{i}(x,Q^2)~,
\label{dc1}
\ee
cf.~\cite{DIC}, where $ n=2,4... $ for $i = NC, +$ and $n=1,3...$ for 
$i = -$, which reads in $x$--space
\be
g_4^i(x,Q^2) = 2xg_5^i(x,Q^2)~.
\label{DIC}
\ee
This relation is not preserved in the presence of target mass corrections
as in the corresponding case for unpolarized deep inelastic scattering
also the Callan--Gross relation~\cite{CG}. 
This is expected since the tensor--structure
in $W_{\mu\nu}$ in the case of $g_5$ and $g_4$ is the same as that
of $F_1$ and $F_2$ for the unpolarized part except the factor of
$S \cdot P$. The breaking term $\Delta_D(x,M^2/Q^2)$,
\be
g_4^i(x,Q^2) = 2xg_5^i(x,Q^2) + \Delta_D\left(x,\frac{M^2}{Q^2}\right)~,
\ee
is of $O(M^2/Q^2)$ and reads
\ba
\Delta_D\left(x,\frac{M^2}{Q^2}\right) &=&  
-  x^2 \frac{d}{dx} x \frac{d^2}{dx^2} \left[
\frac{x^2}{(1+4M^2x^2/Q^2)^{1/2}} \frac{G_2(\xi)}{\xi^2}\right]
\nonumber\\
& & 
- 2 \frac{M^2}{Q^2} x^3 \frac{d^2}{dx^2} \left[\frac{x^2}{(1
+ 4 M^2x^2/Q^2)^{1/2}} \frac{G_2(\xi)}{\xi}\right]
\nonumber\\
& & 
+ 2 x^2 \frac{d}{dx} \left[\frac{x}{(1+4M^2x^2/Q^2)^{1/2}}\frac{G_3(\xi)}
{\xi}\right]~.
\ea
\section{Relations between the Structure Functions~: Twist 3}
\subsection{General Relations}

\vspace{1mm}
\noindent
In the presence of target mass corrections all structure functions 
$\left. g_i\right|_{i=1}^5$ contain twist--3 contributions. On the 
contrary, in the massless limit this is the case for the structure
functions $g_2$ and $g_3$ only~\cite{BK}. The structure functions are 
defined by the hadronic tensor $W_{\mu\nu}$, Eq.~(\ref{had1}). From its 
structure alone it cannot be concluded which structure functions vanish 
in the limit $M \rightarrow 0$. The mass dependence of the scattering
cross sections for longitudinal nucleon polarization, Eq.~(\ref{scaL}),
however, reveals that both the structure functions $g_2$ and $g_3$ do 
only contribute at $O(M^2/Q^2)$. Therefore, a complete account for 
twist--3 contributions requires to consider also the nucleon mass 
corrections for the other polarized structure functions.

Let us consider now the relations between the twist--3 parts of the 
structure functions. From Eqs.~(\ref{g1t3dif}--\ref{g5t3dif}) one derives
the following {\it new} relations between twist--3 parts of different
spin--dependent structure functions~:

\vspace{1mm}
\begin{eqnarray}
 g^i_{1,~\tau=3}(x,Q^2) & = & \frac{4 M^2 x^2}{Q^2}
\left[ g^i_{2,~\tau=3}(x,Q^2)
-2\int_{x}^{1}{\frac{dy}{y} g^i_{2,~\tau=3}(y,Q^2)}\right]~,
\label{t3g1g2}\\
\frac{4 M^2 x^2}{Q^2} g^i_{3,~\tau=3}(x,Q^2) & = &
 g^i_{4,~\tau=3}(x,Q^2)\left(1+\frac{4 M^2 x^2}{Q^2}\right)+
3\int_{x}^{1}{\frac{dy}{y} g^i_{4,~\tau=3}(y,Q^2)}~,
\label{t3g3g4}\\
2 x g^i_{5,~\tau=3}(x,Q^2) &=&
 -\int_{x}^{1}{\frac{dy}{y} g^i_{4,~\tau=3}(y,Q^2)}~.
\label{t3g4g5}
\end{eqnarray}

\vspace{1mm}
\noindent
Here, Eq.~(\ref{t3g1g2}) at the one side and 
Eqs.~(\ref{t3g3g4},\ref{t3g4g5}) on the other side correspond to
different flavor combinations among the twist--3 contributions. This is
similar to the case of the twist--2 terms, where the former case
corresponds to the combination $\Delta q + \Delta\overline{q}$, 
and the latter to $\Delta q - \Delta\overline{q}$.

The corresponding relations for the Mellin moments of the twist--3 part
of the structure functions can be derived from 
Eqs.~(\ref{g1t3NC}--\ref{g5t3-}) and read
\begin{eqnarray}
\int_{0}^{1}dx~x^n g^i_{1,~\tau=3}(x,Q^2) &=& \frac{n+1}{n+3}
\int_{0}^{1}{dx~x^n \frac{4 M^2x^2}{Q^2} g^i_{2,~\tau=3}(x,Q^2)}~,
\label{Mt3g1g2}\\
\int_{0}^{1}{dx~x^n \frac{4 M^2x^2}{Q^2} g^i_{3,~\tau=3}(x,Q^2)} & = &
\frac{n+4}{n+1}\int_{0}^{1}{dx~x^n g^i_{4,~\tau=3}(x,Q^2)} \nonumber\\
&&\hspace{2.5cm} +
\int_{0}^{1}{dx~x^n \frac{4 M^2x^2}{Q^2}
g^i_{4,~\tau=3}(x,Q^2)}~, 
\label{Mt3g3g4}\\
\int_{0}^{1}{dx~x^{n+1} g^i_{5,~\tau=3}(x,Q^2)} &=&
-\frac{1}{2(n+1)}\int_{0}^{1}{dx~x^n g^i_{4,~\tau=3}(x,Q^2)}~.
\label{Mt3g4g5}
\end{eqnarray}
The consistency of these relations is easily checked by partial
integration. Eqs.~(\ref{t3g1g2}--\ref{t3g4g5}) show that the twist--3
contributions to $g_1, g_4$ and $g_5$ vanish in the limit $M \rightarrow
0$. On the other hand, if one keeps terms of $O((M^2/Q^2) \cdot g_{2(3)})$ 
and twist--3 in the scattering cross sections, one has to account also for
the twist--3 terms in $g_1, g_4$ and $g_5$. Because of the arguments given
at the end of section~7 this requires the {\it complete} account for 
target mass corrections.
\subsection{Other Relations}

\vspace{1mm}
\noindent
If the nucleon mass effects are disregarded in the operator product
expansion only the structure functions $g_2$ and $g_3$ receive twist--3
contributions~\cite{BK}. A relation between the matrix elements
$d_n^{q+}$,
$d_n^{q-}$ contributing to $g_2$ and $g_3$, respectively, can be
constructed by {\it assuming} that
\begin{equation}
d_n^{q-} = \left. d_n^{q+}\right|_{\rm val}~.
\label{assum}
\end{equation}
One obtains
\begin{eqnarray}
\label{val1}
\int_0^1 dx x^n \left.
\left[4g_5^-(x,Q^2) - \frac{n+1}{x} g_3^-(x,Q^2)
\right] \right|_{\nu n - \nu p}
&=& 
(n-1) \left[d_n^{u-} - d_n^{d-}\right]~,\\
\label{val2}
\int_0^1 dx x^n \left.
\left[n g_1^{\gamma}(x,Q^2) + (n+1) g_2^{\gamma}(x,Q^2)
\right]\right|_{\gamma p - \gamma n}
&=& 
\frac{n}{12} \left[d_n^{u+} - d_n^{d+}\left. \right]
\right|_{\rm val},~~n = 2,4, \ldots
\end{eqnarray}
Eq.~(\ref{assum}) allows to combine these relations to
\begin{eqnarray}
g_{3, \tau = 3}^{\nu n - \nu p}(x,Q^2) 
= 12 \left[x g_{2, \tau = 3}(x,Q^2)
- \int_x^1 dy g_{2, \tau = 3}(y,Q^2)\right]^{\gamma p - \gamma n}~.
\end{eqnarray}
The target mass corrections break this relation. The correction terms
to Eqs.~(\ref{val1},\ref{val2}) are
\begin{eqnarray}
\int_0^1 dx x^{n} \left[4 g_5^-(x,Q^2) - \frac{n+1}{x} 
g_3^-(x,Q^2)\right]
&=& - (n-1) \sum_q g_V^q g_A^q \sum_{j=0}^{\infty} \left(\frac{M^2}{Q^2}
\right)^j \frac{(n+j)!}{j! (n-1)!} \frac{d_{n+2j}^{q-}}{n+2j}~,\nonumber
\\ \\
\int_0^1 dx x^{n} \left[n g_1^\gamma(x,Q^2) + (n+1) g_2^\gamma(x,Q^2)
\right]
&=&  \sum_q \frac{e_q^2}{4} \sum_{j=0}^{\infty} \left(\frac{M^2}{Q^2}
\right)^j \frac{(n+j)! (n+2j)}{j! (n-1)!} \frac{d_{n+2j}^{q+}}{n+2j}~.
\end{eqnarray}
In general it is not possible to absorb these contributions into the
respective structure functions.

A sum rule  for the first moment of the valence part of $g_1(x,Q^2)$ and 
$g_2(x,Q^2)$ 
\begin{eqnarray}
\int_{0}^{1}{dx x\left[g_1^V(x)+2g_2^V(x)\right]}&=&0.
\label{ELT1}
\end{eqnarray}
was obtained in Ref.~\cite{ELT}. In the zero--mass limit it was shown in 
Ref.~\cite{BK} that this sum rule is formally consistent  with the 
operator product expansion. The valence parts $g_1^V(x)$ and $g_2^V(x)$ 
cannot be isolated for electromagnetic interactions from the complete 
structure functions. One may, however, refer to the charged current case,
Eqs.~(\ref{g1-},\ref{g2-}), and obtains
\begin{eqnarray}
\int_{0}^{1}{dx\, x(g_1^-(x,xQ^2)+2g_2^-(x,Q^2))} &=&
\sum_q{\frac{(g_V^q)^2+(g_A^q)^2}{4}\sum_{j=0}^{\infty}{
\left(\frac{M^2}{Q^2}\right)^j(j+1)d_{2j+1}^{-q} }}~.
\label{ELT2} 
\end{eqnarray}
The l.h.s. of Eq. (\ref{ELT2}) contains only valence quark contributions. 
For individual quark flavors we may write separately
\begin{eqnarray}
\int_{0}^{1}{dx\, x\left[g_1^{V_q}(x,xQ^2)+2g_2^{V_q}(x,Q^2)\right]} 
&=&
\frac{e_q^2}{4}\sum_{j=0}^{\infty}{
\left(\frac{M^2}{Q^2}\right)^j(j+1)d_{2j+1}^{V_q} }~.
\label{ELT2N}
\end{eqnarray}
The matrix elements of the twist--3 operators for odd indices, $d_{2j+1}$,
are proportional to the mass of the
quark. Let us consider the matrix elements 
of the following bi-local operators
\begin{eqnarray}
\Theta^{\beta\mu_1\cdots\mu_n} & = &
{\cal S}_n
\bar q(\gamma_\beta\gamma_5iD^\mu_1\ldots
iD^{\mu_{n}}-\gamma_{\mu_1}\gamma_5iD^\beta\ldots iD^{\mu_{n}})q~,
\label{ELT3}\\
O^{\beta\mu_1\cdots\mu_n} & = &
{\cal S}_n
\bar q i\gamma_5\sigma^{\beta\mu_1}iD^{\mu_2}
\ldots iD^{\mu_{2j}}q~,
\label{ELT4}
\end{eqnarray}
where the symbol ${\cal S}_n$ denotes the symmetrization of the
indices 
$\mu_1,\ldots,\mu_n$. The nucleon matrix elements of these operators are 
related by
\begin{eqnarray}
\langle PS|\Theta^{\beta\mu_1\cdots\mu_n}|PS\rangle & = &
m_q \langle PS|O^{\beta\mu_1\cdots\mu_n}|PS\rangle~.
\label{ELT5}
\end{eqnarray}
In the massive quark case the $\Theta$-- and $O$--operators are not
traceless and their nucleon matrix elements are defined as follows
\begin{eqnarray}
\langle PS|\Theta^{\beta\mu_1\cdots\mu_{2j}}|PS\rangle & =&
{\cal S}_n d_{2j+1}^{V_q}(S^\beta P^{\mu_1}\ldots P^{\mu_{2j}}-
S^{\mu_1}P^{\beta}\ldots P^{\mu_{2j}})+  g_{ij}\,\, {\rm terms}~, 
\label{ELT6}\\
\langle PS|O^{\beta\mu_1\cdots\mu_{2j}}|PS\rangle & =&
{\cal S}_n \frac{a_{2j+1}}{M}(S^\beta P^{\mu_1}\ldots P^{\mu_{2j}}-
S^{\mu_1}P^{\beta}\ldots P^{\mu_{2j}})+  g_{ij}\,\, {\rm terms}~.
\label{ELT7}
\end{eqnarray}
The matrix element in the r.h.s. of Eq.~(\ref{ELT7}) is related to the 
moments of the twist--2 structure function $h_1(x)$, cf. Ref.~\cite{JJ}~,
by 
\begin{equation}
\int_{0}^{1} {dx\, x^{n-1} h_1(x)} =
\int_{0}^{1}{dx\, x^{n-1}\left[h_1(x)-(-1)^{n-1}\overline h_1(x)
\right]} = a_n~.
\label{ELT8}
\end{equation}
Due to the same symmetry properties of operators $\Theta$ and $O$ the 
tensors in the r.h.s. of  the Eqs.~(\ref{ELT6}) and (\ref{ELT7}) are the
same. Therefore, using the expressions for nuclear  matrix elements of 
the above operators and the relation between them, Eq.~(\ref{ELT2N}), the
first moment of $g_1(x)+2g_2(x)$, Eq.~(\ref{ELT2}), can be expressed 
through the moments of transversity distribution
\begin{eqnarray}
\int_{0}^{1}{dx~x\left[g_1(x)+2g_2(x)\right]} &=& \frac{e_q^2}{2}
\frac{m_q}{M} 
\sum_{j=0}^{\infty}{\left(\frac{M^2}{Q^2}\right)^j(j+1)
\int_{0}^{1}{dx~x^{2j} [h_1(x)-\overline{h}_1(x)]}}~.
\label{ELT9}
\end{eqnarray}
After resummation one obtains
\begin{eqnarray}
\int_{0}^{1}{dx x \left[g_1(x)+2g_2(x)\right]} 
&=& \frac{e_q^2}{2}\frac{m_q}{M} 
\int_{0}^{1}{dx \frac{h(x)-\overline{h}(x)}{\ds
\left(1-\frac{M^2 x^2}{Q^2}\right)^2}}~.
\label{ELT10}
\end{eqnarray}
For $Q^2 > M^2$, which was implicitly assumed in performing the light
cone expansion above, the integral in the r.h.s. of Eq.~(\ref{ELT10}) is 
finite. Therefore the r.h.s. of Eq.~(\ref{ELT10}) vanishes in the limit 
$m_q \rightarrow 0$. This is equivalent to $d_{2j+1}^{V_q} \rightarrow 0$
and Eq.~(\ref{ELT1}) is found to be formally consistent with the
the result of the operator product expansion.
\section{Conclusions}

\vspace{2mm}
\noindent
We have calculated the target mass corrections for all polarized structure 
functions for both neutral and charged current deep inelastic scattering
in the case of conserved currents. The results were obtained by using the
local light cone expansion of the Compton amplitude for forward
scattering. In the evaluation we used the approach of Ref.~\cite{GP}.
The target mass corrections imply besides the twist--2 terms twist--3 
contributions for all polarized structure functions. Only after the 
inclusion of the target mass corrections the description of the polarized
scattering cross sections can be regarded as being completed at the 
twist--3 level, since the contributions to $g_1, g_4$ and $g_5$ are of the
same order as those to $g_2$ and $g_3$ discussed previously. The 
corrections were both represented in terms of the integer moments which 
result from the light cone expansion and their analytic continuation and 
Mellin inversion to $x$-space. The latter representations are directly
applicable in experimental analyses. 

We investigated the effect of the target mass corrections on the sum rules
connecting the polarized structure functions in lowest order in the
coupling constant. For the twist--2 contributions both the 
Wandzura--Wilczek relation~\cite{WW} and the relation derived in 
Ref.~\cite{BK} are preserved, whereas the Dicus relation~\cite{DIC},
similarly to the Callan--Gross relation~\cite{CG} in the unpolarized
case, receives a correction. It was also shown that the Wandzura--Wilczek
relation is preserved in the presence of quark--mass corrections. A 
previously derived integral relation between the twist--3 valence 
contributions to $g_2$ and $g_3$~\cite{BK} receives target mass 
corrections. A relation for the first moment of a combination between 
$g_1$ and $g_2$~\cite{ELT} is preserved.

Three new integral relations were derived for the twist--3 contributions 
of the polarized structure functions. They hold without further assumption
on the flavor combinations of the related structure functions. In the case
of the relation between the twist--3 contributions to $g_1$ and $g_2$ 
experimental test can be performed in the foreseeable future by precise
measurements of the longitudinal and transversely polarized deep inelastic
$eN$--scattering cross sections at lower values of $Q^2$.

\vspace*{2mm}
\noindent
{\bf Acknowledgement.}\\
We would like to thank W.L. van Neerven for discussions. The work was 
supported in part by EU contract FMRX-CT98-0194(DG 12 - MIHT) and the 
Alexander-von-Humboldt Foundation.

\newpage
\section{Appendix~:\\
The quark mass contribution to the Wandzura-Wilczek relation}

\vspace{2mm}
\noindent
We consider the parts of forward Compton scattering amplitude
$\widehat T_{\mu\nu,spin}^i$, Eqs.~(\ref{TCompP}) and (\ref{TCompM}),
which correspond to the structure functions $g_1(x)$ and $g_2(x)$ in the 
massive quarks case,\footnote{The resummation of quark mass effects
was considered for the unpolarized structure functions in 
Refs.~\cite{GP,QUARK}.}
\begin{eqnarray}
\widehat T_{1}^{\pm}&\sim& 
\varepsilon_{\mu\alpha\nu\beta}q^\alpha 
\sum_{n~even,odd}{q^{\mu_1}\cdots q^{\mu_n}
 \Bigl(\frac{2}{Q^2-m_X^2+m_q^2}\Bigr)^{n+1}
\Theta^{\pm\beta\{\mu_1\cdots\mu_n\}}}~.
\end{eqnarray}
Here $m_X$ is the mass of the struck quark and $m_q$ the final quark mass.
The initial--state quark is not necessarily on shell, but we assume that
its wave function is the usual free-particle Dirac spinor for a particle 
with mass $m_X$.

In the massive quark case the $\Theta$--operators are not traceless 
anymore and have no definite twist. To construct the  nucleon matrix 
elements for these operators, we have to decompose them into traceless 
ones. The most general form for a such decomposition is
\begin{eqnarray}
\Theta^{\pm\beta\{\mu_1\cdots\mu_n\}}& = & \sum_{j=0}^{\infty}{
A(n,j) {\cal O}^{\pm\beta\{\mu_1\cdots\mu_{n-2j}\}} 
\underbrace{g\ldots g}_{j}~(m_X^2)^j }
\nonumber \\
&+&
\sum_{j=0}^{\infty}
{B(n,j){\cal O}^{\pm\mu_{\alpha}\{\beta\mu_1\cdots\mu_{n-2j-1}\}}
\underbrace{g\cdots g}_{j}~(m_X^2)^j}
 \nonumber \\
&+&
\sum_{j=0}^{\infty}
{C(n,j){\cal O}^{\pm\mu_\alpha\{\mu_1\cdots\mu_{n-2j}\}}
\underbrace{g\ldots g}_{j} g_{\beta\mu_\beta}~(m_X^2)^j}~.
\label{theta}
\end{eqnarray}
Because of the
antisymmetric tensor $\varepsilon_{\mu\alpha\nu\beta}$ the 
third term $\propto C(n,j)$ does not contribute to $\widehat T^{\pm}_1$. 
Symmetrizing the other two operators in Eq.~(\ref{theta}) we separate 
the twist--2 part in the forward Compton scattering amplitude. The 
symmetric part of the remaining two operators are identical. Therefore
one  obtains
\begin{eqnarray}
\widehat T_{1}^{\pm}&\sim& \varepsilon_{\mu\alpha\nu\beta}q^\alpha 
\sum_{n~even,odd}{\sum_{j=0}^{[n/2]}{ \left(\frac{2}{Q^2-m_X^2+m_q^2}
\right)^{n+1}
\left[A(n,j)+B(n,j)\right]}}\nonumber\\
&& \times
q^{\mu_1}\cdots q^{\mu_n}
\Theta^{\pm\beta\{\mu_1\cdots\mu_{n-2j}\}}\underbrace{g\ldots g}_{j} (m_X^2)^j
\nonumber\\
&=& \varepsilon_{\mu\alpha\nu\beta}q^\alpha 
\sum_{n~even,odd}{\sum_{j=0}^{[n/2]}{
\left(\frac{Q^2}{Q^2-m_X^2+m_q^2}\right)^{n+1}
\left(\frac{2}{Q^2}\right)^{n-2j+1}\left(\frac{-4 m_X^2}{Q^2}\right)^{j}}}
\nonumber\\
&&\times
\left[A(n,j)+B(n,j)\right]
q^{\mu_1}\cdots q^{\mu_{n-2j}}\Theta^{\pm\beta\{\mu_1\cdots\mu_{n-2j}\}} 
\nonumber\\
&=& \varepsilon_{\mu\alpha\nu\beta}q^\alpha 
\sum_{l~~even,odd}{\left(\frac{2}{Q^2-m_X^2+m_q^2}\right)^{l+1}
X_l\left[\frac{4m_X^2Q^2}{(Q^2-m_X^2+m_q^2)^2}\right]}
\nonumber\\
&&\times
q^{\mu_1}\cdots q^{\mu_l}\Theta^{\pm\beta\{\mu_1\cdots\mu_l\}}~.
\label{T1Fin}
\end{eqnarray}
In the last step we changed the summation index from $n$ to $l=n-2j$ and 
introduced the function $X_l(z)$,
\begin{eqnarray}
X_l(z) &=& \sum_{j=0}{ (-z)^j \left[A(l+2j,j)+B(l+2j,j)\right]}~.
\end{eqnarray}
As can be seen from Eq.~(\ref{T1Fin}) we obtain the same tensor structure
as in Eq.~(\ref{T1forWW}) which leads to the Wandzura--Wilczek relation. 
Because the dependence on the quark masses is the same in the structure
functions $g_1(x)$ and $g_2(x)$ the Wandzura--Wilczek relation is not 
violated by these corrections.
\newpage


\begin{thebibliography}{999}
%
\bibitem{LC}
K.G. Wilson, Phys. Rev. {\bf 179} (1969) 1699;\\
R.A. Brandt and G. Preparata, Fortschr. Phys. {\bf 18} (1970) 249;\\
W. Zimmermann, in: {\sf Elementary Particle Physics and Quantum Field
Theory}, Brandeis Summer Inst., Vol.~1, (MIT Press, Cambridge, 1970),
p.~397;\\
Y. Frishman, Ann. Phys. {\bf 66} (1971) 373;
Phys. Rep. {\bf C13} (1974) 1.
%
\bibitem{SLAC}
A. Bodek et al., Phys. Lett. {\bf 30} (1973) 1084;
Phys. Rev. {\bf D21} (1979) 1471;\\
J.S. Poucher et al., Phys Rev. Lett. {\bf 32} (1974) 118;\\
E.M. Riordan et al., Phys. Rev. Lett. {\bf 33} (1974) 561;\\
S. Stein et al., Phys. Rev. {\bf D11} (1975) 1884;\\
W.B. Atwood et al., Phys. Lett. {\bf B64} (1976) 479;\\
M.D. Mestayer et al., Phys. Rev. {\bf D27} (1983) 285;\\
R.G. Arnold et al., Phys. Rev. Lett. {\bf 52} (1984) 727;\\
S. Dasu et al., Phys. Rev. Lett. {\bf 61} (1988) 1061;\\
R.E. Taylor, Rev. Mod Phys. {\bf 63} (1991) 573;\\
H.W. Kendall, Rev. Mod. Phys. {\bf 63} (1991) 597;\\
J.I. Friedmann, Rev. Mod. Phys. {\bf 63} (1991) 615;\\
L.W. Whitlow et al., Phys. Lett. {\bf B282} (1992) 475; SLAC--report--357
(1990) and references therein; \\
W. Bartel et al., Phys. Lett. {\bf 28B} (1968) 148;\\
W. Albrecht et al., Nucl. Phys. {\bf B13} (1969) 1;\\
J. Moritz et al., Nucl. Phys. {\bf B41} (1972) 336;\\
E. Lohrmann, in~:
Proc. Lund International Conf. on Elementary Particles, ed.
G. von Dardel, Lund, Sweden, June 1969, (Berlingska Bocktryckeriet, 1969),
 p.~13.
%
\bibitem{POLE}
D. Adams et al., SMC collaboration, Phys. Rev. {\bf D56} (1997) 5330;\\
B. Adeva et al., SMC collaboration, Phys. Rev. {\bf D58} (1998) 112001;\\
K. Ackerstaff et al., HERMES collaboration, Phys. Lett. {\bf B404} (1997)
383;\\
A. Airapetian et al., HERMES collaboration, {\tt hep-ex/9807015}.
%
\bibitem{G2EXP}
K. Abe et al., E143 collaboration, Phys. Rev. Lett. {\bf 76} (1996) 587;
Phys. Rev. {\bf D58} (1998) 112003;\\
K. Abe et al., E154 collaboration, Phys. Lett. {\bf B404} (1997) 377.
%
\bibitem{HT1}
S. Gottlieb, Nucl. Phys. {\bf B139} (1978) 125; PhD thesis
{\sf A New Twist in Deep Inelastic Scattering}, Princeton University, 
1978.
%
\bibitem{HT2}
S. Wada, Progr. Theor. Phys. {\bf 62} (1979) 475; Nucl. Phys. {\bf B202}
(1983) 201;\\
M. Okawa, Nucl. Phys. {\bf B172} (1980) 481; {\bf B187} (1981) 71;\\
H.D. Politzer, Nucl. Phys. {\bf B172} (1980) 349;\\
E.V. Shuryak and A.I. Vainstein, Phys. Lett. {\bf B105} (1981) 65;
Nucl. Phys. {\bf B199} (1982) 451; {\bf B201} (1982) 141;\\
R.L. Jaffe and M. Soldate, Phys. Lett. {\bf B105} (1981) 467;
Phys. Rev. {\bf D26} (1982) 49;\\
S.P. Lutterell, S. Wada, and B.R. Webber, Nucl. Phys. {\bf B188} (1981)
219;\\
S.P. Lutterell and S. Wada, Nucl. Phys. {\bf B197} (1982) 290;\\
R.K. Ellis, W. Furmanski, and R. Petronzio, Nucl. Phys. {\bf B207} (1982)
1; {\bf B212} (1983) 29;\\
C.S. Lam and M.A. Walton, Can. J. Phys. {\bf 63} (1985) 1042;\\
A.P. Bukhvostov  and G.V. Frolov, Sov. J. Nucl. Phys. {\bf 45} (1987)
704;\\
J.W. Qui, Phys. Rev. {\bf D42} (1990) 30.
%
\bibitem{HT3}
J. Bl\"umlein and W.L. van Neerven, to appear.
%
\bibitem{GP}
H.~Georgi and H.D.~Politzer, Phys.~Rev. {\bf D14} (1976) 1829.
%
\bibitem{PR}
A.~Piccione and G.~Ridolfi, Nucl.~Phys. {\bf B513} (1998) 301.
%
\bibitem{Nachtmann}
O.~Nachtmann, Nucl.~Phys. {\bf B63} (1973) 237.
%
\bibitem{XI}
V. Baluni and E. Eichten, Phys. Rev. Lett. {\bf 37} (1976) 1181; Phys.
Rev. {\bf D14} (1976) 3045;\\
A. De R\'ujula, H. Georgi, and H.D. Politzer, Ann. Phys. (NY) {\bf 103}
(1977) 315;\\
D.J.~Gross, S.B.~Treiman, and F.A.~Wilczek, Phys.~Rev. {\bf D15} (1977) 
2486;\\
A.~De R\'ujula, H.~Georgi, and H.D.~Politzer, Phys.~Rev. {\bf D15} (1977) 
2495.
%
\bibitem{WA}
S.~Wandzura, Nucl.~Phys. {\bf B122} (1977) 412.
%
\bibitem{MU}
S. Matsuda and T. Uematsu, Nucl. Phys. {\bf B168} (1980) 181.
%
\bibitem{KU}
H. Kawamura and T. Uematsu, Phys. Lett. {\bf B343} (1995) 346.
%
\bibitem{BK}
J.~Bl\"umlein and N.~Kochelev, Phys. Lett. {\bf B381} (1996) 296;
Nucl.~Phys. {\bf B498} (1997) 285.
%
\bibitem{DIC}
D.A.~Dicus, Phys. Rev. {\bf D5} (1972) 1367.
%
\bibitem{WW}
S.~Wandzura and F.~Wilczek, Phys. Lett. {\bf B72} (1977) 195.
%
\bibitem{Ji}
X.~Ji, Nucl. Phys. {\bf B402} (1993) 217.
%
\bibitem{FRANKF}
M.~Maul, B.~Ehrnsperger, E.~Stein, A.~Sch\"afer,
Z.~Phys. {\bf A356} (1997) 443.
%
\bibitem{GS}
A.H.~Guth and D.E.~Soper, Phys.Rev. {\bf D12} (1975) 1143.
%
\bibitem{CARLS}
For a uniqueness theorem  see, E. Carlson, {\sf Thesis}, Univ. Uppsala,
1914;\\
E.C. Titchmarsh, {\sf Theory of Functions}, (Oxford University Press,
Oxford, 1939), Chapt.~9.5.
%
\bibitem{GKP}
R.L. Graham, D.E. Knuth, and O. Patashnik, {\sf Concrete Mathematics},
2nd edition, (Addison--Wesley, Reading/MA, 1994).
%
\bibitem{MSG}
J.L.~Miramontes and J.~S\'anchez Guill\'en, Z.~Phys. {\bf C41} (1988) 247.
%
\bibitem{BJT}
K.~Bitar, P.W.~Johnson, and W.-K.~Tung, Phys.~Lett. {\bf B83} (1979) 114.
%
\bibitem{DDOR}
A.~Devoto, D.W.~Duke, J.F.~Owens, and R.G.~Roberts, Phys.~Rev. {\bf D27} (1983) 508.
%
\bibitem{EMP1}
F. Eisele, M. Gl\"uck, E. Hoffmann, and E. Reya, Phys. Rev. {\bf D26}
(1982) 41;\\
R.M. Barnett, Phys. Rev. Lett. {\bf 48} (1982) 1657; 
Phys. Rev. {\bf D27} (1983) 98;\\
R.G. Arnold et al., Phys. Rev. Lett. {\bf 52} (1984) 727;\\
J.F. Gunion, P. Nason, and R. Blankenbecler, Phys. Rev. {\bf D29} (1984)
2491.
%
\bibitem{EMP2}
K. Varvell et al., BEBC collaboration, Z. Phys. {\bf C36} (1987) 1.
%
\bibitem{CN}
J.M. Cornwall and R.E. Norton, Phys. Rev. {\bf 177} (1969) 2584.
%
\bibitem{BC}
H.~Burkhardt and W.N.~Cottingham, Ann. Physics (New York) {\bf 56} (1970)
453.
%
\bibitem{CG}
C.G.~Callan and D.J.~Gross, Phys. Rev. Lett. {\bf 22} (1969) 156.
%
\bibitem{ELT}
A.V. Efremov, O.V. Teryaev and E. Leader, Phys.~Rev. {\bf D55} (1997) 
4307.
%
\bibitem{JJ}
R.L.~Jaffe and X.~Ji, Nucl.~Phys. {\bf B375} (1992) 527.
%
\bibitem{QUARK}
R.M. Barnett, Phys. Rev. {\bf D14} (1976) 70;\\
R.K. Ellis, G. Parisi, R. Petronzio, Phys. Lett. {\bf B64} (1976) 97;\\
R. Barbieri, J. Ellis, M.K. Gaillard, and G.G. Ross, Phys. Lett. 
{\bf B64} (1976) 171; Nucl. Phys. {\bf B117} (1976) 80;\\
R. Brock, Phys. Rev. Lett. {\bf 44} (1980) 1027.
\end{thebibliography}
\end{document}